 \documentclass[final,5p,times]{elsarticle}

\usepackage{lineno}
\usepackage{url}
\usepackage{graphicx}
\usepackage{amstext,amsmath,amssymb,amsthm}
\usepackage{upgreek}
\usepackage{enumerate}
\usepackage{txfonts}
\usepackage{graphicx}
\usepackage{subfigure}
\usepackage{multirow}
\usepackage{multicol}
\usepackage[T1]{fontenc}

\begin{document}

\begin{frontmatter}



\title{In-flight calibration of the \emph{Insight-Hard X-ray Modulation Telescope}}


\author[IHEP]{Xiaobo Li*}
\author[IHEP]{Xufang Li}
\author[IHEP]{Ying Tan}
\author[IHEP]{Yanji Yang}
\author[IHEP]{Mingyu Ge}
\author[IHEP]{Juan Zhang}
\author[IHEP]{Youli Tuo}
\author[IHEP]{Baiyang Wu}
\author[IHEP]{Jinyuan Liao}
\author[IHEP]{Yifei Zhang}
\author[IHEP,UCAS]{Liming Song}
\author[IHEP]{Shu Zhang}
\author[IHEP,UCAS]{Jinlu Qu}
\author[IHEP,UCAS]{Shuang-nan Zhang}
\author[IHEP,UCAS]{Fangjun Lu}
\author[IHEP]{Yupeng Xu}
\author[IHEP]{Congzhan Liu}
\author[IHEP]{Xuelei Cao}
\author[IHEP]{Yong Chen}
\author[IHEP]{Jianyin Nie}
\author[IHEP]{Haisheng Zhao}
\author[IHEP]{Chengkui Li}

\cortext[cor1]{\emph{Email address: lixb@ihep.ac.cn (Xiaobo Li)}}
\address[IHEP]{Key Laboratory of Particle Astrophysics, Institute of High Energy Physics, Chinese Academy of Sciences, Beijing, China}
\address[UCAS]{University of Chinese Academy of Sciences, Beijing, China}

\begin{abstract}
 We present the calibration of the \emph{Insight-Hard X-ray Modulation Telescope} (\emph{Insight-HXMT}) X-ray satellite, which can be used to perform timing and spectral studies of bright X-ray sources.  \emph{Insight-HXMT} carries three main payloads onboard: the High Energy X-ray telescope (HE), the Medium Energy X-ray telescope (ME) and the low Energy X-ray telescope (LE). In orbit, the radioactive sources, activated lines, the fluorescence lines and celestial sources are used to calibrate the energy scale and energy resolution of the payloads. The Crab nebular is adopted as the primary effective area calibrator and empirical functions are constructed to modify the simulated effective areas of the three payloads respectively. The systematic errors of HE, compared to the model of the Crab nebular, are less than 2\% in 28--120\,keV and  2\%--10\% above 120\,keV. The systematic errors of ME are less than 1.5\% in 10--35\,keV. The systematic errors of LE are less than 1\% in 1--7\,keV except the Si K--edge (1.839\,keV, up to 1.5\%) and less than 2\% in 7--10\,keV.

\end{abstract}


\end{frontmatter}


\section{Introduction}
\label{sec:1 introduction}
As China's first X-ray astronomical satellite, the \emph{Insight-Hard X-ray Modulation Telescope} (\emph{Insight-HXMT}, \citep{Zsn2019},\citep{LuFangjun..HXMT..2018},\citep{S.Zhang..2018}) was successfully launched on June 15th, 2017 into a low earth orbit with an altitude of 550\,km and an inclination of 43\,degrees. As an X-ray astronomical satellite with a broad band in 1--250\,keV, three payloads are designed onboard \emph{Insight-HXMT}, i.e., High Energy X-ray telescope (HE) using 18 NaI(Tl)/CsI(Na) phoswich scintillation detectors for 20--250\,keV band\citep{Lcz2019}, Medium Energy X-ray telescope (ME) using 1728 Si-PIN detectors for 5--30\,keV band\citep{Cxl2019}, and Low Energy X-ray telescope (LE) using 96 SCD detectors for 1--15\,keV band\citep{Cy2019}. The three payloads are installed on a same supporting structure to achieve the same pointing direction, thus they can simultaneously observe the same source. They all have collimators to confine different kinds of field of view (FOV). The background of the large FOV detectors is more complicated than that of small FOV detectors, so we focus on the calibration of the small FOV detectors for ME and LE.

HE can be used to study a wide variety of timing phenomena due to the its large area and microsecond time-tagging. Although the background level of HE is high at about 500\,mCrab, it can be used to study the spectra of bright sources without the influence of pile-up and dead-time. Compared with other types of CCD detectors, the readout of LE is very fast and the pile-up effect can be ignored even the flux reaches to about 8\,Crab. LE also performs well for studying the spectra of bright sources.

 Ground calibration experiments and modeling of the response matrix of the payloads were performed before launch. In this paper, we describe the in-flight refinement of the calibration of \emph{Insight-HXMT}. Section 2 gives an overview of the instruments. Section 3 describes the energy scale and energy resolution calibration, including the description of the parameters that are supplied to the three instruments. Section 4 describes the response matrix calibration and we also briefly describe the spectral response prior to launch using the calibration data on ground. Section 5 describes the calibration of effective areas based on the Crab nebular observations and simultaneous observations between \emph{NuSTAR} and \emph{Insight-HXMT} to verify the calibration of the effective areas.  Section 6 describes the systematic errors of the three instruments onboard \emph{Insight-HXMT} which can be used in the spectral fitting.
The response of the collimators, or the Point Spread Function (PSF) of \emph{Insight-HXMT} is described else where \citep{Nang2019}.
The absolute timing accuracy of \emph{Insight-HXMT} is about 50\,us from the time of arrival (TOA) of the Crab Pulsar compared with observations of other telescopes, which has also been described else where \citep{XB.LiSPIE2018}.

\section{Instruments description} \label{sec:2 instrument description}
\begin{figure}
	\centering
		\includegraphics[width=0.44\textwidth]{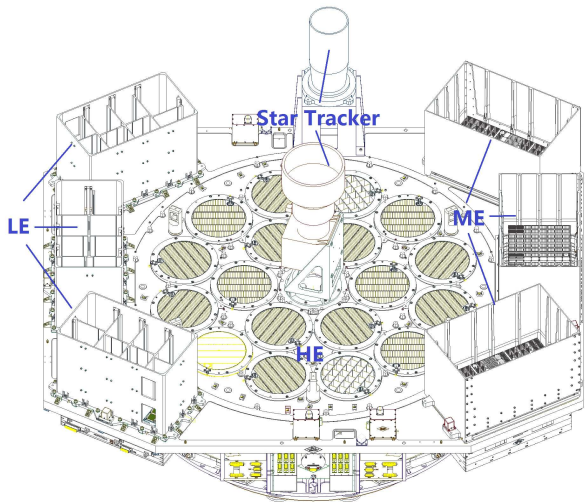}
	\caption{The layout and orientations of the \emph{Insight-HXMT} payloads. The 18 phoswich detectors of HE with collimators are in the middle of the picture. The FOVs are defined by the collimators. The three boxes on the lef are the LE telescope and the three boxes on the right are the ME telescope. }
	\label{FIG:HXMT}
\end{figure}

The three instruments onboard \emph{Insight-HXMT} are\\
 slat-collimated type of telescopes as shown in Figure \ref{FIG:HXMT}. The rectangular metallic grid collimators are installed on top of the detectors to shield the photons outside of the FOVs.
The collimators also have different orientations with a step of 60 degrees\citep{Zsn2019}. Except the large and small FOVs, the three instruments also have blind FOVs to estimate the background.

\subsection{HE} \label{sec:2.1 HE instrument}

HE adopts an array of 18 NaI(Tl)/CsI(Na) phoswich as the main detectors\citep{Lcz2019}. The diameter of each phoswich is 190\,mm. The thicknesses of NaI(Tl) and CsI(Na) are about 3.5\,mm and 40\,mm, respectively. The working temperature is actively controlled at $18\pm2{}^{\circ}\mathrm{C}$.
 The incident X-ray with most of its energy deposited in NaI(Tl) is regarded as a NaI(Tl) event. CsI(Na) is used as an active shielding detector to reject the background events from backside and events with partial energy loss in the NaI(Tl). The scintillation photons generated within the two crystals can be collected by the same photomultiplier tube (PMT). Signals from the PMT are pulse shaped to distinguish NaI(Tl) events and CsI(Na) events because of the different decay time in the two crystals. The energy loss, time of arrival and pulse width of each detected event are measured, digitized and telemetered to the ground. The CsI(Na) can also  be used as a gamma-ray burst (GRB) monitor. The detected energy range in high gain mode is about 50--800\,keV and is changed to 250\,keV--3\,MeV in low gain mode for CsI(Na) if the high voltage of PMT is decreased. In this paper, we only describe the calibration of NaI(Tl) events in high gain mode. The dead time of each detector is recorded online every second and also telemetered to the ground.

For each phoswich detector, a radioactive source $^{\mathrm 241}$Am with an activity of 200\,Bq is embedded into a plastic scintillator and viewed by a separate Multi-Pixel Photon Counter\citep{Lcz2019}. They are all mounted in the collimator and used as an automatic gain control (AGC) detector. A coincident measurement between the AGC detector\,($\alpha$ particle of 5.5\,MeV) and phoswich detector\,(X-ray of 59.5\,keV) is labeled as a calibration event. The calibration events are saved like norm events, but with a different flag. The response of NaI(Tl) is not uniform in its large surface, so the calibration events are just used as the gain control and are not suitable to calibrate the energy scale of NaI(Tl) detectors in-orbit.
 Table \ref{tab:tab1 HXMT character} summarizes the characteristics of  HE.


\begin{table}[htb]\footnotesize
  \centering
  \caption{Properties of HXMT detectors}\label{tab:tab1 HXMT character}
  \resizebox{0.5\textwidth}{!}{
    \tabcolsep 0.01in
    \begin{tabular}{cccccc}
    \hline
    Characteristic & HE  & ME & LE \\
    \hline
  Energy range (keV) & 28-250 &10-35 & 1-10 \\
  Energy resolution & 18$\%$@60\,keV & 13.6$\%$@22\,keV &  1.5$\%$@6.4\,keV\\
  Time resolution (us) & 2 & 6.4 & 10\\
  FOV & $1.1^{\circ} \times 5.7^{\circ}$   & $1^{\circ} \times 4^{\circ}$  &  $1.6^{\circ} \times 6^{\circ}$    \\
      & $5.7^{\circ} \times 5.7^{\circ}$  & $4^{\circ} \times 4^{\circ}$ &$4^{\circ} \times 6^{\circ}$ \\
  Detector & NaI(Tl), CsI(Na) & Si-PIN &  SCD \\
  Open area (${\rm cm}^{2}$) & 4270 & 850 & 300 \\
  Operating temperature & $18\pm2{}^{\circ}\mathrm{C}$ &$-40^{\circ}\mathrm{C}$$\thicksim$$-10^{\circ}\mathrm{C}$ & $-75^{\circ}\mathrm{C}$$\thicksim$$-40^{\circ}\mathrm{C}$ \\
        \hline
    \end{tabular}
    }
\end{table}

\subsection{ME} \label{sec:2.2 ME instrument}
ME consists of three detector boxes and each box has 576 Si-PIN detector pixels read out by 18 ASIC (Application Specified Integrated Circuit)\citep{Cxl2019}. Each ASIC is responsible for the readout of 32 pixels.  The thickness of the Si-PIN is 1\,mm. The energy loss and arrival time of each detected event are measured, digitized and telemetered to the ground.
 In each detector box, two $^{\mathrm 241}$Am sources are fixed in the corner of two ASIC and each one illuminates four pixels\citep{Cxl2019}.
 The working temperature of Si-PIN detectors in orbit is between $-40^{\circ}\mathrm{C}$ and $-10^{\circ}\mathrm{C}$ .
 Table \ref{tab:tab1 HXMT character} also summarizes the characteristics of  ME.

Si-PIN detectors are fixed on the ceramic chip by silver glue. When the energy of incident X-rays is greater than 25.5 keV (K-edge of Ag), they have some probability of penetrating the Si-PIN and interact directly with silver. Fluorescence lines of silver will be generated due to the photoelectric effect with electrons in K-shell of silver and escape from the silver glue and then are detected by the Si-PIN detectors.

Although the Si-PIN signals are already screened by the onboard threshold, the events transmitted to the ground still contain low-energy thermal and electrical noise component, which varies significantly in orbit. To remove these noise events, a higher threshold must be applied to ME detectors. The new thresholds range from 6 to 10\,keV for different Si-PIN detectors and we choose the new threshold at 10\,keV for all of them in spectral analysis.

\subsection{LE} \label{sec:2.3 LE instrument}

LE also consists of three detector boxes and each box contains 32 CCD236 which is a kind of Swept Charge Devices (SCD)\citep{Cy2019}. CCD236 is the second-generation SCD, which has been developed by e2v company. It is built with a sensitive area of about 4 ${\rm cm}^{2}$ of each CCD236 and has four quadrants. Each quadrant has 100*100 pixels. The L-shaped readout mechanism has fast signal response but the position information is lost\citep{Cy2019}. The readout cycle is 100\,us. For events with energy above the onboard threshold, the energy loss and the readout time of each detected event are measured, digitized and telemetered to the ground. Besides this, LE also has the forced trigger events, which record the amplitude of the noise or the pedestal offset for each CCD detector every 32\,ms\citep{Cy2019}. The forced trigger events are also saved like physical events, but with a different type.

The spreading of the charge cloud over several pixels may cause split events. These split events may be read out in adjacent readout periods. In this paper, we focus on the calibration of the single events in order to reduce the effect of the charged particles.
The working temperature in-orbit for CCD236 is between $-75^{\circ}\mathrm{C}$ and $-40^{\circ}\mathrm{C}$. The characteristics of LE is also summarized in Table \ref{tab:tab1 HXMT character}.

\section{Energy scale and resolution calibration} \label{sec:3 energy scale and resolution cal}
The information of energy scale and resolution of each detector should be included in the generation of its response matrix. The following subsections are dedicated to describing the details of in-flight energy scale and resolution calibration of \emph{Insight-HXMT}.
\subsection{Energy to channel model of HE} \label{sec:3.1 cal of he gain}
Many experiments have indicated that the light output response of the NaI(Tl) crystal is not proportional to the deposited energies of X-rays and electrons \citep{NaI..nRP..1998},\citep{Khodyuk..nPR..2010}. The light output is also non-monotonic at iodine K-edge at 33.17\,keV and 50.2\,keV for HE detectors \citep{XF.LiHE2019}.  On ground, three quadratic functions at three energy ranges (below iodine K-edge, from iodine K-edge to 50.2\,keV and above the 50.2\,keV)  were used to parameterize the energy-channel (E-C) relationship of HE detectors\citep{XF.LiHE2019}.

\begin{figure}
	\centering
		\includegraphics[width=0.44\textwidth]{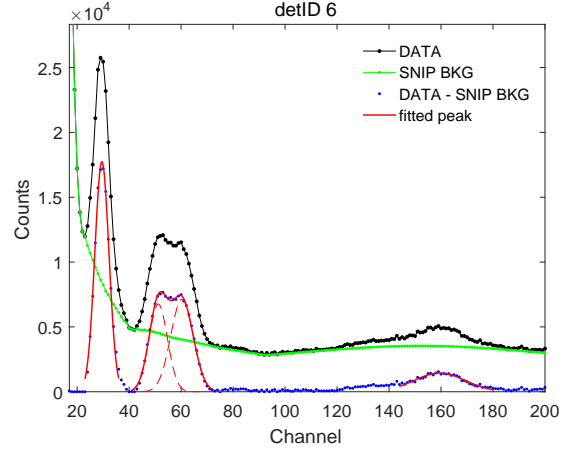}
	\caption{The process of illustrating how to obtain the peak centroids of the four lines in the spectrum of blank sky observation. The black point data are the measured spectrum of one detector (detID=6). The green line is the continuum obtained by the SNIP algorithm; The blue is the continuum subtracted spectrum with only line profiles; The red lines are the Gaussian functions that are used to fit the peak centroids of the four lines.}
	\label{FIG:0}
\end{figure}

The background of HE in-orbit is dominated by internal activation effects. Prominent background lines due to activation of iodine by cosmic and SAA protons are evident at 31, 56, 67 and 191\,keV \citep{XB.LiSPIE2018}.
These four lines could be used to calibrate the E-C relation or energy scale of HE.
 Statistics sensitive Nonlinear Iterative Peak clipping (SNIP,\citep{SNIP2016}) method, which is also widely used in the gamma-ray spectrometry, is utilized to determine the continuum of the observed background spectrum.
 After subtracting the continuum component, the peak centroids of the four lines can be fitted with Gaussian functions.
  As shown in Figure \ref{FIG:0}, it illustrates the process of how to obtain the peak centroids of the four lines in the spectrum of blank sky observation.

\begin{figure}
	\centering
\includegraphics[width=0.44\textwidth]{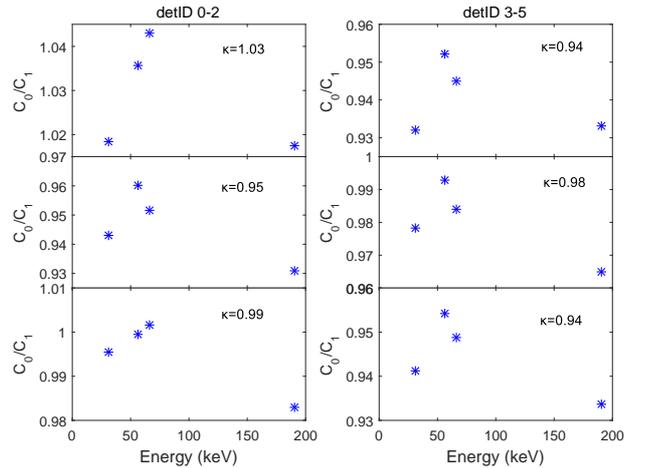}
	\caption{The peak ratios of the net channels between pre-launch and post-launch for the four background lines (31, 56, 67, 191\,keV) measured by detID 0-5. $C_{0}$ represents the net peak channel on ground and $C_{1}$  represents that in-orbit. }
	\label{FIG:1}
\end{figure}

  According to the E-C relationship on ground, the peak centroids of the four lines are different with the expected values. As shown in Figure \ref{FIG:1}, the peak ratios of the net channels (where the offset of the electronics has been subtracted) between pre-launch and post-launch of the four background lines are not equal to 1 and here we just show the ratios for detID 0-5.
  Therefore, the E-C of HE detectors in orbit are different with the pre-launch calibration results, and thus must be re-calibrated.

In pre-launch calibration experiments, to model the E-C relation \citep{XF.LiHE2019}, we defined
 \begin{equation}\label{equ:heecunderground}
   E=A_{1}x^{2}+A_{2}x +A_{3},
 \end{equation}
 where x is the net channel of the full energy peak, $E$ is the energy of incident photons from the monochromatic radiation, $A_{1}, A_{2}$ and $A_{3}$ are the fitting parameters.
 In-flight, the parameter $\kappa$ that lessens the difference among the four background lines could be adopted to correct the pre-launch E-C and $\kappa$ is affected by the efficiency of the light collection and high voltage of the PMT. We construct the following model to represent the in-flight EC,
 \begin{equation}\label{equ:heecinorbit}
   E=A_{1}(\kappa x)^{2}+A_{2}(\kappa x) +A_{3},
 \end{equation}
 where $\kappa$ is the ratio of net channels between pre-launch and orbit. The half of the maximum and minimum of $\kappa$ of the four background lines as plotted in Figure \ref{FIG:1} is regarded as the ratio of different detectors of HE. The accuracy of this gain calibration is estimated to be about 1\% based on the deviations of the four calibration points.
 In order to validate the energy scale results, we jointly fitted the energy of cyclotron resonance scattering features (CRSF) of Her X-1 using the simultaneous observation with \emph{NuSTAR} telescope and obtained consistent results at around 37.5\,keV \citep{GC.Xiao2019}.

\begin{figure}
	\centering
\includegraphics[width=0.45\textwidth]{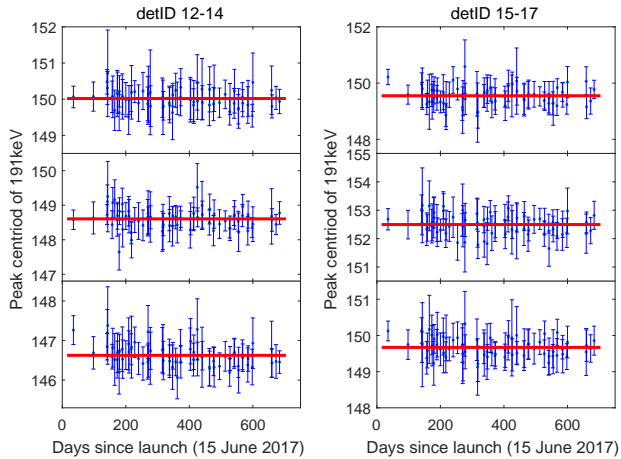}
	\caption{The peak centroid of 191\,keV background line against time. The left panel is for detID 12, 13, 14 and the right panel is for detID 15, 16, 17. The standard deviations of most detectors are at 0.3 channel. }
	\label{FIG:2}
\end{figure}

\begin{figure}
	\centering
\includegraphics[width=0.46\textwidth]{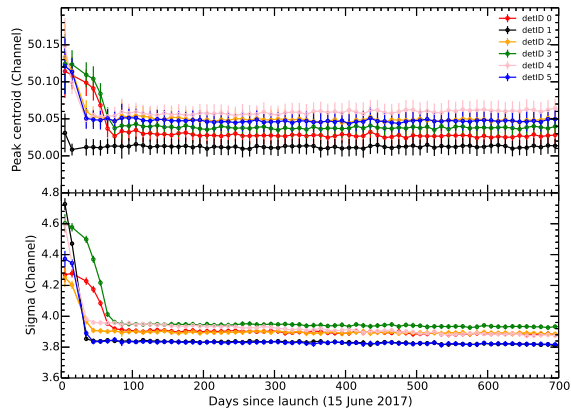}
	\caption{The top panel shows the peak centroid of 59.5\,keV from $^{\mathrm 241}$Am versus time for detID 0-5. The bottom panel shows the sigma of 59.5\,keV from $^{\mathrm 241}$Am against time for detID 0-5. After about 90 days in-orbit, the peak centroids and sigma are stable at 0.01 channel and 0.02 channel, respectively.}
	\label{FIG:3}
\end{figure}

In order to monitor the energy scale of HE, we make full use of regular blank sky observations which are used to estimate the background model. Figure \ref{FIG:2} shows the peak centriods of 191\,keV line versus time. The standard deviations of the peak centroids range from 0.23 to 0.49 channel for all the HE detectors and most of the detectors are at 0.3 channel. We also utilize the data of $^{\mathrm 241}$Am, which is located in the edge of each detector and permanently illuminates a small area of one HE detector, to monitor the energy scale. The top panel of Figure \ref{FIG:3} shows the variations of the peak centroid of the 59.5\,keV versus time. The variations are less than 0.01 channel for all the detectors after about 90 days.
From Figure \ref{FIG:2} and the top panel of Figure \ref{FIG:3}, the energy scale of HE remains stable after about three months in-orbit.

\subsection{Energy resolution model of HE} \label{sec:3.2 cal of he resolution}

In the pre-launch calibration experiments of HE, we modeled the resolution of HE in channel space as
 \begin{equation}\label{equ:heresunderground}
   R(x)=\frac{B_{1}+B_{2}x+B_{3}\sqrt{x}}{x},
 \end{equation}
 where $R(x)$ is the Full Width at Half Maximum (FWHM) in channel space and $x$ is the channel value, $B_{1}, B_{2}$ and $B_{3}$ are the fitting parameters \citep{XF.LiHE2019}.
 In-orbit, we have introduced another parameter $\lambda$ to estimate the resolution in channel space as
  \begin{equation}\label{equ:heresorbit}
   R(x)=\frac{B_{1}\lambda+B_{2}x+B_{3}\sqrt{x\lambda}}{x}.
 \end{equation}

\begin{figure}
	\centering
\includegraphics[width=0.45\textwidth]{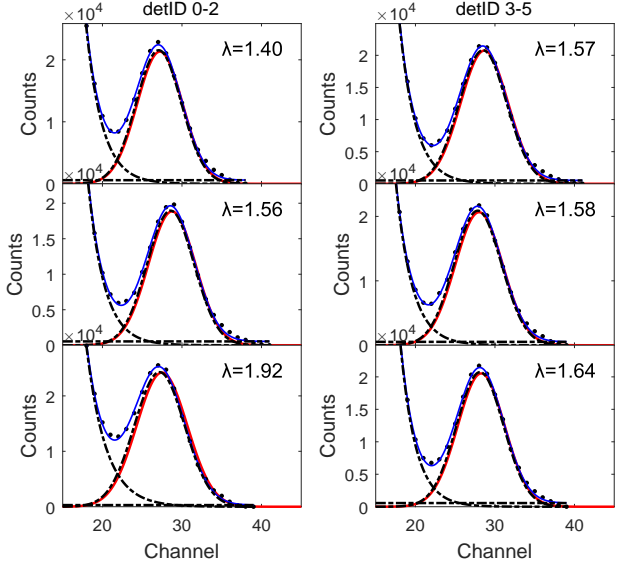}
	\caption{The process of determining the parameter $\lambda$. The black point data are the measured blank sky spectrum near 31\,keV. The three black dashed lines represent the three functions, Gaussian, Exponent and Constant respectively. The blue line is the sum of three functions and the red line is the simulated profile of 31\,keV.}
	\label{FIG:4}
\end{figure}

Before launch, we utilized an object-oriented toolkit, Geant4 (\citep{Geant4..2003}, \citep{Geant4..2006}, \citep{Geant4..2016}, version 4.9.4), to estimate the background of HE in orbit and found some lines can contribute to the profile of 31\,keV line as shown in Table \ref{tab:tab2 31keV lines}. The energies and the corresponding intensities which come from the Geant4 simulation, are also displayed in Table \ref{tab:tab2 31keV lines}.
 We use the in-flight E-C to get the peak channel of the blended lines and adjust the parameter $\lambda$ from 0.9 to 3 with bisection method to broaden the lines. Till the difference between the width of simulated profile of 31\,keV and that of the observed is less than 0.001 channel, the parameter $\lambda$ of each detector can be obtained. This process is shown in Figure \ref{FIG:4}.
 A Gaussian function with an exponential function and a small constant provides an excellent fit to the profile of 31\,keV by use of the blank sky data. The width of 191\,keV is also used to verify the parameter $\lambda$.

\begin{table}[htb]\footnotesize
  \centering
  \caption{The energies and intensities of the mixed 31\,keV line.}\label{tab:tab2 31keV lines}
          \begin{tabular}{|c|c|c|}
    \hline
    No. & Energy(keV)  & Intensity(\%) \\
    \hline
     1 & 25.25 &1.31  \\
    \hline
   2 & 27.75& 2.98  \\
   \hline
   3 & 28.25& 1.14  \\
   \hline
   4 & 29.25& 8.82  \\
   \hline
   5 & 30.25& 25.44  \\
   \hline
   6 & 30.55& 0.93  \\
   \hline
   7 & 31.75& 42.00  \\
    \hline
   8 & 33.25& 10.77  \\
   \hline
   9 & 36.25& 1.45  \\
   \hline
   10 & 39.75& 2.35  \\
   \hline
   11 & 40.25 & 2.80  \\
   \hline
    \end{tabular}

\end{table}

Except background line of 31\,keV, we also monitor the width of 59.5\,keV emitted by the $^{\mathrm 241}$Am. A Gaussian function is used to fit the full energy peak profile of 59.5\,keV, the sigma of the Gaussian versus time is shown in the bottom panel of Figure \ref{FIG:3}. The variations are less than 0.02 channel after about 90 days in orbit.
The resolution of the 18 HE detectors also remains stable after about 90 days launch.

\subsection{Energy scale and resolution calibration of ME} \label{sec:3.3 cal of ME gain and res}

\begin{figure}
	\centering
	\includegraphics[width=0.48\textwidth]{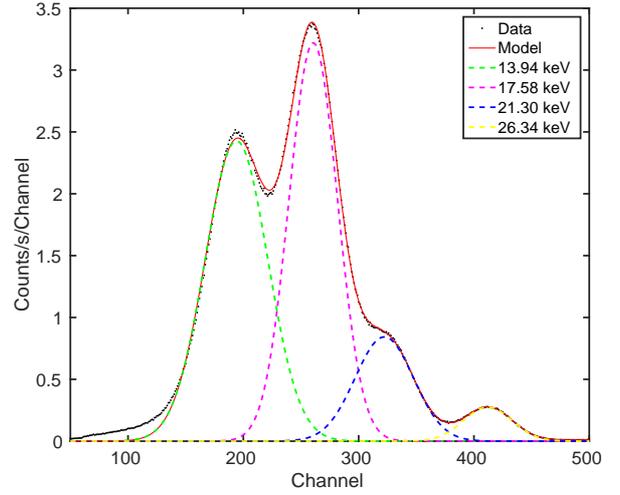}
	\caption{Data and Gaussian fits to the $^{\mathrm 241}$Am nuclear lines collected in the blank sky observations. The red line is the sum of the four Gaussian functions with dashed lines.}
	\label{FIG:6_3}
\end{figure}

Each ME detector box contains two $^{\mathrm 241}$Am sources which can continuously illuminate 8 pixels.
 The spectrum of pixels carried with $^{\mathrm 241}$Am can be accumulated during the blank sky observations as shown in Figure \ref{FIG:6_3}.
The energies and intensities about each line are from the database of National Nuclear Data Center (NNDC, \citep{NNDC}.)  After taking the efficiencies and resolution of ME detectors into account, the model energies of the four lines can be determined as
13.94\,keV, 17.58\,keV, 21.30\,keV, and 26.34\,keV.

In the calibration experiments on ground, the E-C relation of each pixel was linear from 11 keV to 30 keV using the spectrum of $^{\mathrm 241}$Am source. The slopes and intercepts of E-C for all pixels were not a constant at different temperatures \citep{XB.LiSPIE2018}. The slopes of E-C increased with the temperature whereas the intercepts remained almost the same at different temperatures.
The slopes and intercepts at seven temperatures ranging from  $-30^{\circ}\mathrm{C}$ to $-2^{\circ}\mathrm{C}$ were measured on ground and stored in CALDB as the primary E-C calculation.
The slopes and intercepts at a given temperature can be obtained from the linear interpolation at two adjacent temperatures for each pixel.

\begin{figure}
	\centering
	\includegraphics[width=0.48\textwidth]{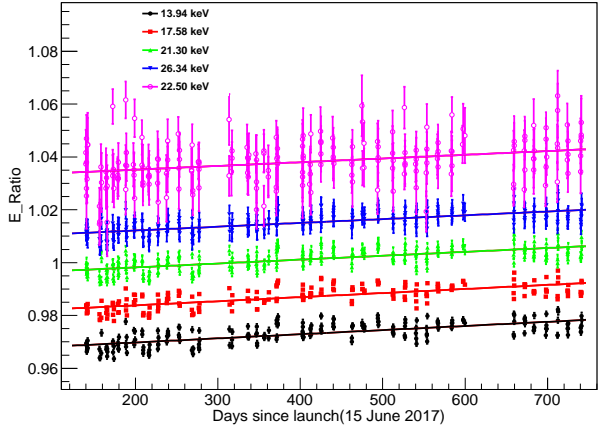}
	\caption{The fit of the energy peak (with 1$\sigma$ error bars) to the four lines of $^{\mathrm 241}$Am and Ag line in detector box 0 versus time. The energy peak has scaled with a different constant.  A linear fit is used to describe the evolution of the five lines.}
	\label{FIG:6}
\end{figure}

In orbit, we use the pre-launch E-C model of each pixel to convert primary ADC channel to energy, then utilize the same linear function of energy versus pulse invariant (PI) channel to get the PI channel spectrum of $^{\mathrm 241}$Am as x-axis plotted in Figure \ref{FIG:6_3}.
 A model with four Gaussians provides an excellent fit to the $^{\mathrm 241}$Am spectrum between 11\,keV and 27\,keV.
 The spectrum of other pixels not illuminated with $^{\mathrm 241}$Am  are also generated from the blank sky data to fit the Ag peak (22.5\,keV) using the pre-launch E-C.
 A model with a power-law and a Gaussian line is used to fit the data near Ag peak between 20\,keV and 24\,keV. Over this bandpass, this simple model provides an acceptable fit.
  The peak values of the five lines increase slowly with time as shown in Figure \ref{FIG:6}.  A linear function is used to fit the evolution of the five lines. Till now,  they increase by less than 1\%.
 In the current status of software (HXMTDAS V2.02), this effect is not corrected in the gain for this small and slow change. But in the next software update, this effect will be added.

\begin{figure}
	\centering
\includegraphics[width=0.47\textwidth]{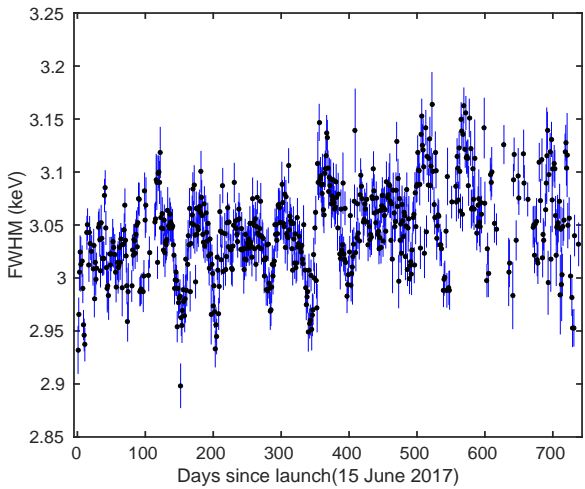}
	\caption{The FWHM fit (with 1$\sigma$ error bars) to the $^{\mathrm 241}$Am line in detector box 0 versus time. The resolution has increased by 0.5\% till 700 days after launch.}
	\label{FIG:6_2}
\end{figure}

The FWHM of Ag line and the four nuclear lines from $^{\mathrm 241}$Am are utilized to monitor the resolution of ME in orbit.
 As shown in Figure \ref{FIG:6_2},  the FWHM of $^{\mathrm 241}$Am, which is used as the same parameter for the four lines in the spectral fitting, is plotted as a function of time.
 Although, there are some structures due to the effect of temperature, the change is very small.
 The resolution at 13.94\,keV has increased by 0.5\% till 700 days in orbit.
 Therefore, the resolution measured on ground can also be used for the in orbit data for all the pixels.

\subsection{Energy scale calibration of LE} \label{sec:3.3 cal of LE gain}
For LE, the data suitable for parameterizing the energy scale and monitoring its variations come from three regular observed sources: the internal background lines (Ni, Cu, and Zn),  observations of Cas A with rich lines (Si, S, Fe, et al) and regular monitoring observations of the Crab Nebula providing an opportunity to measure the location of the K-edge of silicon at 1.839\,keV.

\begin{figure}
	\centering
\includegraphics[angle=-90,width=0.46\textwidth]{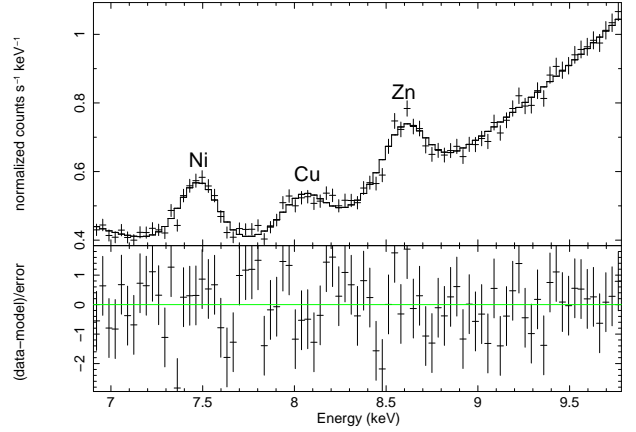}
	\caption{The observed spectrum of blank sky measured by LE. The background lines of Ni, Cu, Zn are prominent in the measured spectrum. The black line in the top panel is the model to fit the energy peak of the three lines. The bottom panel shows the residuals of the fit.}
	\label{FIG:7}
\end{figure}

The supernova remnant Cassiopeia A produces several strong lines easily visible in the LE spectrum \citep{Zsn2019}. Cas A is observed almost every month as a calibration source if the solar avoidance angle is allowed. We have fitted a model with a power-law continuum and several Gaussian lines to the data between different ranges as listed in Table \ref{tab:tab3 LEcasAfitrange} to obtain the energy peaks of these lines using the pre-launch E-C calibration. The emission lines of Ni, Cu and Zn generated by materials near the CCD detectors during the blank sky observations are also used to calibrate the energy gain of LE. As shown in Figure \ref{FIG:7}. the three lines are prominent in the measured spectrum of blank sky.

\begin{table}[htb]\footnotesize
  \centering
  \caption{Energy ranges of different lines for fitting.}\label{tab:tab3 LEcasAfitrange}
          \begin{tabular}{|c|c|c|}
    \hline
    Lines & Energy(keV)  & Fit Range(keV) \\
    \hline
     Mg, Al & 1.355, 1.475 &1.20--1.60  \\
    \hline
    Si & 1.861 & 1.62--2.04  \\
   \hline
    S& 2.456& 2.12--2.74  \\
   \hline
    S,Ar & 2.899, 3.133 & 2.69--3.31  \\
   \hline
   Ca & 3.892& 3.55--4.21  \\
   \hline
   Fe & 6.637 & 6.32--7.01  \\
    \hline
   Ni,Cu,Zn & 7.472, 8.041, 8.631 & 6.91--9.55 \\
   \hline
    \end{tabular}

\end{table}

In the calibration experiments of LE on ground, the E-C relation of each CCD was also linear. The slopes and intercepts of E-C for CCD were not constant at different temperatures \citep{XB.LiSPIE2018}.  The same method for ME was also used to obtain the slopes and intercepts at different temperatures. In orbit, the forced trigger events of LE are accumulated every minute to fit the peak of the pedestal, then the physical events in this minute will subtract the pedestal value to obtain the pure pulse height. As for LE, the channel is referred to as the pedestal subtracted one.
We use the pre-launch E-C results at different temperatures stored in CALDB to fit the peak energies of the lines listed in Table \ref{tab:tab3 LEcasAfitrange} with all the data of in-flight observations of Cas A and blank sky.

First, we check the effect of temperature and the differences between different CCDs in each observed spectrum of Cas A.
 The differences of Si and S energies between different small FOV CCDs is less than 8\,eV at different observation time.
So the data of small FOV CCDs are used to accumulate the spectrum of Cas A lines.

\begin{figure}
	\centering
\includegraphics[width=0.45\textwidth]{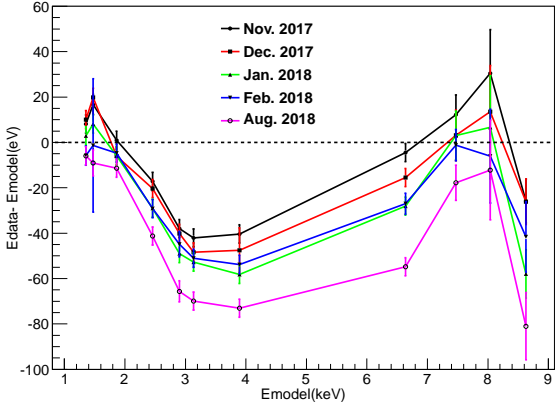}
	\caption{The difference of the energy between data and model for Cas A lines. The data energy is fitted by spectrum used the pre-launch EC of LE. Different color line means the fit result on different observation date.}
	\label{FIG:8}
\end{figure}

\begin{figure}
	\centering
\includegraphics[width=0.45\textwidth]{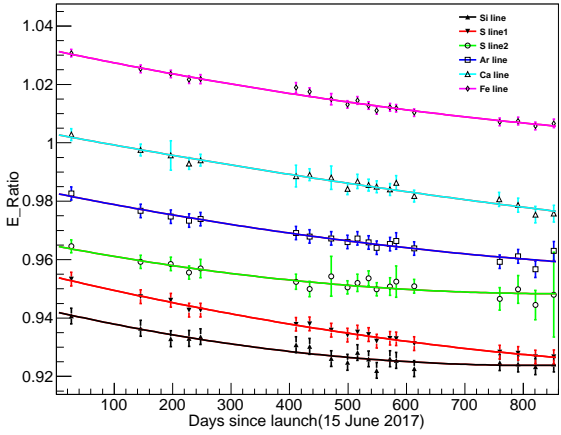}
	\caption{Energy fit (with 1$\sigma$ error bars) to Cas A. The peak of the lines decreased with time. A quadratic polynomial fit is used to describe the evolution.}
	\label{FIG:9}
\end{figure}

The observed spectrum of \emph{XMM/MOS} is also used to fit the energies and intrinsic width of Cas A lines together with LE spectrum at different time. The fit results of energy peaks by \emph{XMM/MOS} are regarded as the model energies of the lines as shown in the middle column of Table \ref{tab:tab3 LEcasAfitrange}.
If the pre-launch E-C of LE is suitable for the in-flight data, the observed energy minus the model energy of the lines would be expected to be around zero.
 But the gain has shifted in a non-linear way as shown in Figure \ref{FIG:8} on different observation date.
 From Figure \ref{FIG:9},  all the peaks of Cas A decrease gradually and this phenomenon may be caused by the decrease of the charge transfer efficiency of the CCDs.
In order to describe the evolution, a quadratic polynomial function is used to fit the change. From the fit result as shown in Figure \ref{FIG:9}, the peak values could be obtained on any day even the Cas A is invisible due to the observation constraint.

\begin{figure}
	\centering
\includegraphics[width=0.45\textwidth]{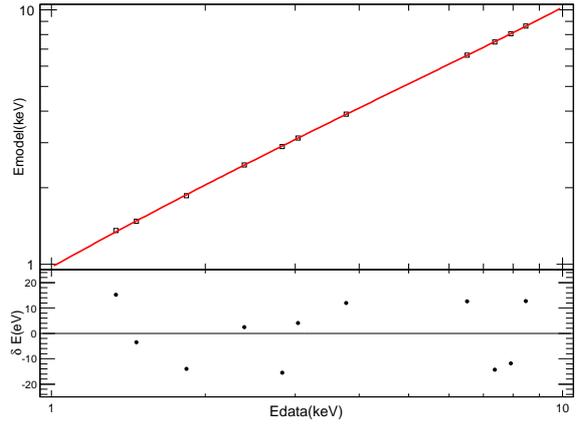}
	\caption{A cubic function is used to fit the in-flight energy gain of LE. The x-axis is the line energy from the evolution fit as shown in Figure \ref{FIG:9}. The y-axis is the corresponding model energy from \emph{XMM/MOS}.  The fit errors of the energy peaks are small and thus ignored. The bottom panel shows the residuals of the fit result.}
	\label{FIG:10legain}
\end{figure}

 According to the residual shape shown in Figure \ref{FIG:8}, a cubic function is used to describe the gain model of LE as shown in Figure \ref{FIG:10legain}. The x-axis is the energy from the polynomial fit result of different peaks as show in Figure \ref{FIG:9} on the same day and y-axis is the model energy fit with \emph{XMM/MOS}.
We compare the cubic function on different days and decide to supply the gain of LE every month to get the accurate gain results. The parameters of the cubic function in each month are stored in CALDB.

\begin{figure}
	\centering
		\includegraphics[width=0.45\textwidth]{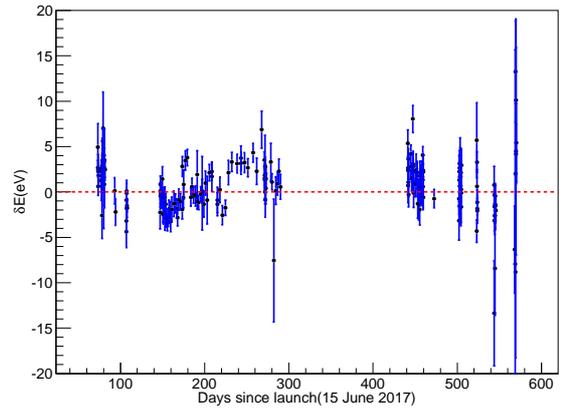}
	\caption{The fit result of Si K-edge minus the K-edge energy at 1.839\,keV versus time. Most of the differences are within
 $\pm 5$\,eV. }
	\label{FIG:10}
\end{figure}

We also use the the following model to fit the K-edge of silicon to verify the gain of LE when Crab is observed,
 \begin{equation}\label{equ:LEedge}
   Y = A_{0}\,{\rm {erfc}}(\frac{E-E_{0}}{\sigma_{0}}) + C_{0},
 \end{equation}
where $E_{0}$ is the fit energy of Si K-edge, $A_{0}$ is the normalization coefficient, $C_{0}$ is the constant term and $\sigma_{0}$ is the width of $E_{0}$. $\rm {erfc}$ is the complementary error function. $Y$ is the rate of Cas A spectrum.
The differences between the fit energy and K-edge energy at 1.839\,keV are shown in Figure \ref{FIG:10} and most of them are within
 $\pm 5$\,eV. Therefore, the gain calibration of LE is effective.

\subsection{Energy resolution calibration of LE} \label{sec:3.3 cal of LE resolution}
After the calibration of LE gain, we generate the Cas A spectrum using the new gain in-orbit again.
 The widths of Si, S, and Fe using the pre-launch response file of LE are jointly fitted with \emph{XMM/MOS} using Xspec. If the energy resolution of LE keeps the same as the measurement on ground, the intrinsic width of Si, S, and Fe will be same as the fit result of \emph{XMM/MOS}.
 Actually, the fitted intrinsic widths of Si, S, and Fe for LE are larger than that of \emph{XMM/MOS}, so the energy resolution is also changed, compared with the pre-launch calibration results.
 After subtracting the intrinsic width of Si, S, and Fe from the fit result of \emph{XMM/MOS}, the extra broadening of the resolution of LE is obtained and it evolves with time and temperature.
 In order to parameterise the extra broadening of LE, we define the following two dimensional function,
  \begin{equation}\label{equ:lefwhm2d}
   W(t,T)= c_{0}+c_{1}t+c_{2}T+c_{3}t^{2}+c_{4}T^{2}+c_{5}tT,
 \end{equation}
where $W(t,T)$ is the extra broadening of LE except the pre-launch spreading,  $t$ is the observation time and $T$ is the temperature, $c_{0}-c_{5}$ are the fitting parameters.

\begin{figure}
	\centering
\includegraphics[width=0.45\textwidth]{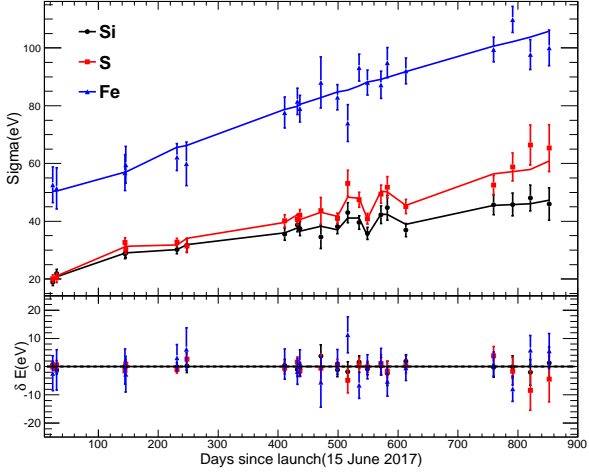}
	\caption{Extra broadening width of Si, S, and Fe plotted against the time at different observation temperature. The model described in Equation (\ref{equ:lefwhm2d}) is also plotted as the lines. In the lower panel, the residuals of the fit are plotted against time for the Si, S, and Fe respectively. The differences between data and model are less than 10\,eV.}
	\label{FIG:11}
\end{figure}

\begin{figure}
	\centering
\includegraphics[width=0.45\textwidth]{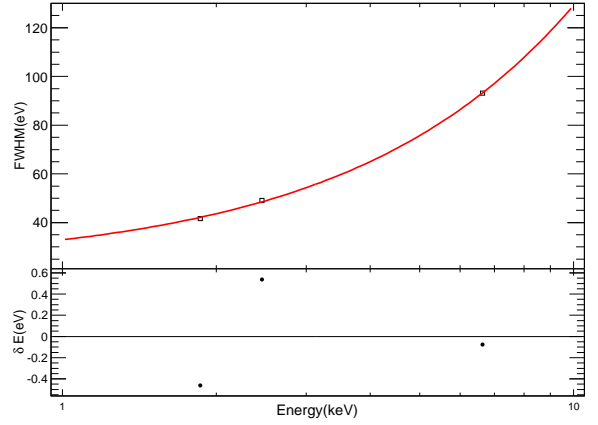}
	\caption{A linear function is used to describe the the extra energy broadening of LE at different energies. The fit errors of the extra broadening are small and thus ignored. The bottom panel shows the fit residuals.}
	\label{FIG:12}
\end{figure}

 The results are plotted for each line in Figure \ref{FIG:11}. The residuals for the Si, S, and Fe are plotted in the bottom panel of Figure \ref{FIG:11}. For an observation, we calculate the mean temperature in its good time interval (GTI) and also the mean time. From the Equation (\ref{equ:lefwhm2d}),  the extra broadening of Si, S, and Fe can be derived for this observation.
 Once the extra boarding widths of Si, S, and Fe have been derived, a liner function is used to fit the extra broadening of LE as shown in Figure \ref{FIG:12}.

\section{Response matrix calibration} \label{RMF cal}
\subsection{HE} \label{sec:4.1 rmf of HE}
Extensive pre-launch calibration experiments and modeling of the response were performed for HE detectors \citep{XF.LiHE2019}. About 32 discrete energies covering the band from 20\,keV to 356\,keV were used to calibrate the energy scale and resolution of HE detectors.
The 32 energies and their corresponding full peak channels were used to get the non-proportional response (NPR) of NaI(Tl) to electrons.

Geant4 (version 4.10.2) is utilized to perform the simulations. The model of low energy electromagnetic interaction, G4EmLivermorePhysics, is invoked in our simulation, and the fluorescence line and Auger process are activated. The primary events, including the type, energy,
direction, and position of the initial particles are generated according to the calibration experiment setup on ground.
Geant4 treats the particles one by one and tracks the trajectories of primary particles and secondary particles step by step.

\begin{figure}
	\centering
\includegraphics[width=0.39\textwidth]{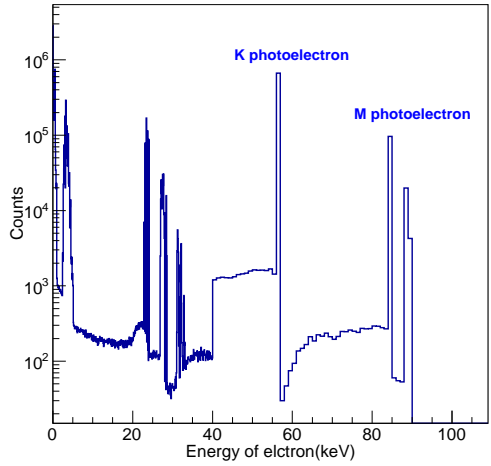}
	\caption{Energy distribution of Auger electrons, photoelectrons and Compton scattered electrons liberated in NaI(Tl) following the absorption of 90\,keV X-ray photons. The photoelectrons generated in the K, M and L of Iodine shell are shown.}
	\label{FIG:13}
\end{figure}

At each step, we record the information of particles, particle type, kinetic energy, deposited energy and physical process involved and so on. Then the recorded information of every step can be used to select the particles of interest.
The following model is used to represent the NPR of NaI(Tl) crystal to electrons,
 \begin{gather}
  \label{equ:henpr}
   NPR(E_{\rm{e}})=p_{1}x^{4}+p_{2}x^{3}+p_{3}x^{2}+p_{4}x+p_{5},\\
   x=\lg{(E_{\rm{e}})},
 \end{gather}
where $E_{\rm{e}}$ is the kinetic energy of electrons generated by the incident X-ray
interacting with NaI(Tl) crystal through the phot-electric effect and Compton scattering.
Figure \ref{FIG:13} is an example showing the simulated distribution of $E_{\rm{e}}$, which is generated in NaI(Tl) by absorption of 90\,keV X-ray photons.
$NPR(E_{e})$ is the pulse height generated by electrons with energy of $E_{\rm{e}}$ per keV with unit of channel/keV.
$p_{1}-p_{5}$ are the fitting parameters.

\begin{figure}
	\centering
		\includegraphics[width=0.45\textwidth]{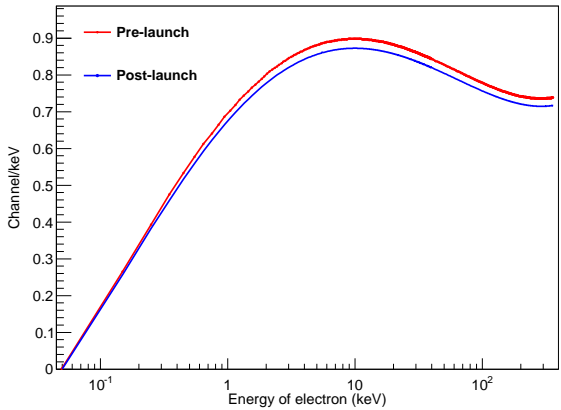}
	\caption{The pre-launch and post-launch NPR model of HE one detector (detID=0). }
	\label{FIG:14npr}
\end{figure}

\begin{figure}
	\centering
		\includegraphics[width=0.45\textwidth]{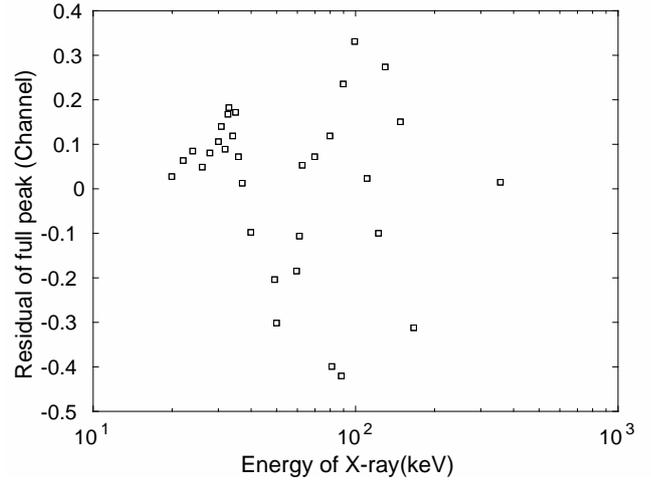}
	\caption{The residuals of full-energy peak of a detector (detID=0) between the NPR model and data for 32 energies measured on ground. }
	\label{FIG:14fullpeakRes}
\end{figure}

We convolve the distribution of kinetic energy of electrons with NRP model to get the theoretical channel of full-energy peak.
A minimization method called fmincon in MATLAB \citep{matlab} is used to find the minimum of the nonlinear multi-variable function. After minimization, the parameters of NPR can be derived and the NPR function is shown in Figure \ref{FIG:14npr}.
The NPR model on ground and in-orbit are both plotted in Figure \ref{FIG:14npr} using the different input of E-C relation.
Before launch, we compared the differences of full peak channel of 32 energies between measured and predicted with NPR model; most of the differences are within $\pm 0.2$ channel as plotted in Figure \ref{FIG:14fullpeakRes}.

\begin{figure}[htb]
  \centering
  \subfigure[]{
     \includegraphics[width=0.35\textwidth]{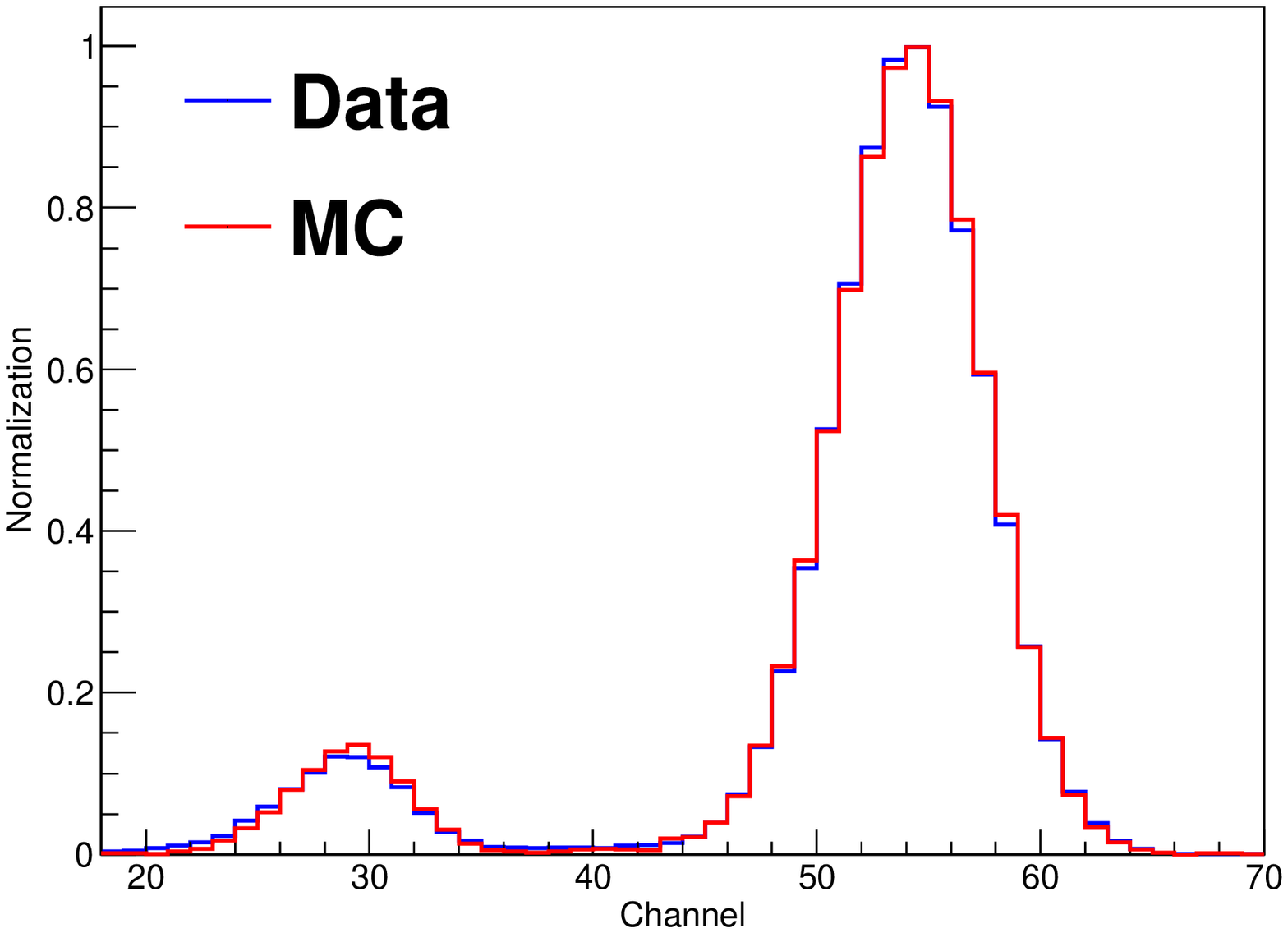}}
  \subfigure[]{
      \includegraphics[width=0.35\textwidth]{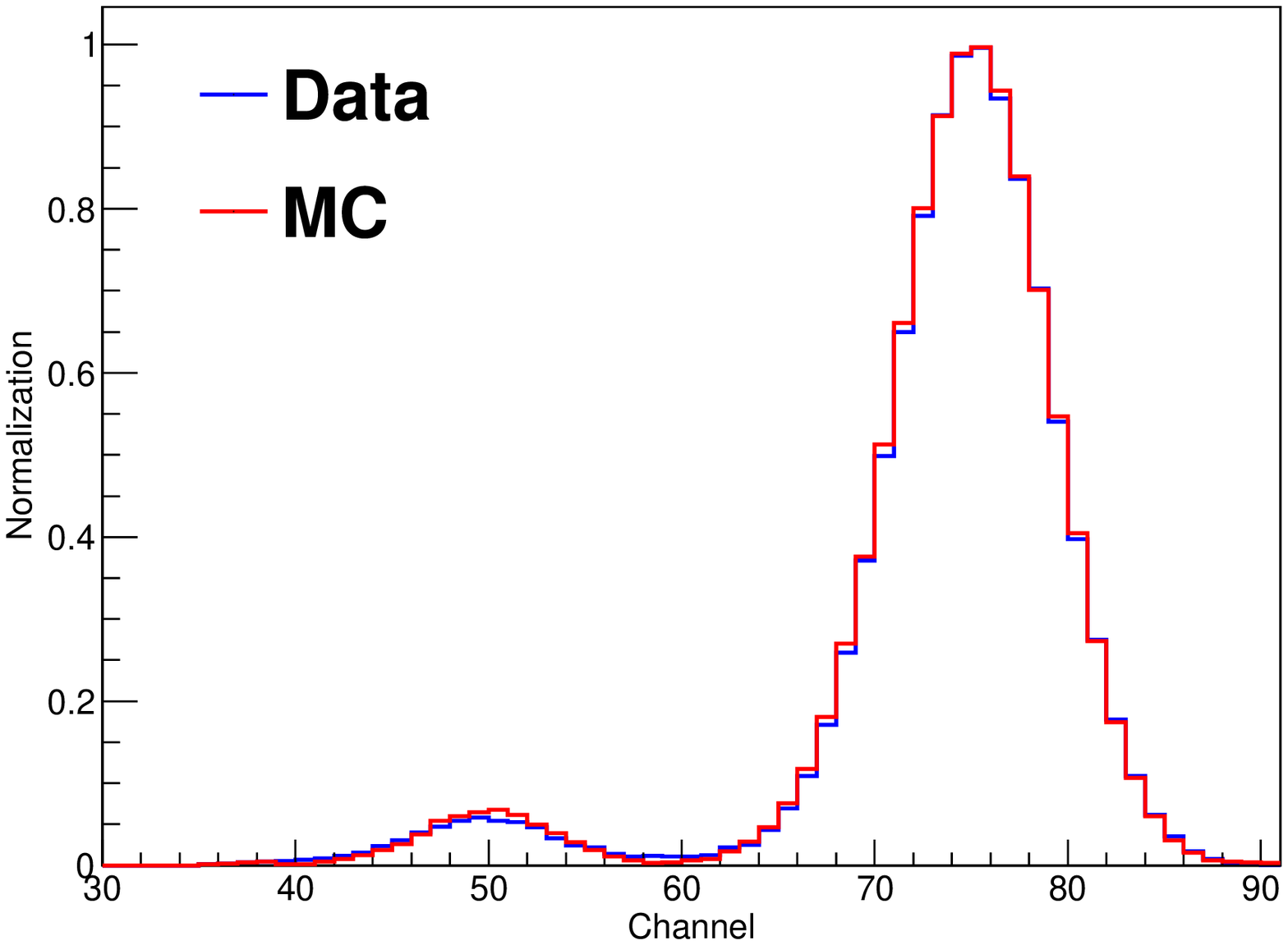}}
    \caption{(a) Simulated spectrum and measured spectrum from monochromatic radiation with energy of 62\,keV. (b) Simulated spectrum and measured spectrum from monochromatic radiation with energy of 90\,keV. The x-axis is the measured channel of HE and y-axis is the normalized counts.}
  \label{FIG:15}
\end{figure}

 The pre-launch energy resolution in channel space was used to spread the simulated spectrum and we also compared it with the measured spectrum as shown in Figure \ref{FIG:15}.  The centroids of full energy peak and escape peak are the same as the data.
 As a result, the NPR model can predict the response function very well. If we do not take the NPR of electrons in NaI(Tl) in the Geant4 simulation into account and only use the deposited energy to get the simulated spectrum, the escape peak shifts by about 3\,keV in the data when the incident photons have energy of 90\,keV.
 Once we have verified the simulated response with the measured 32 energies on ground, we generate the response function at other energies.

 In orbit, the E-C model and resolution in channel space can be utilized to produce the NPR model and response matrix files as same as we have done on ground.

\subsection{ME} \label{sec:4.2 rmf of ME}
 The pre-launch RMFs was calibrated across the ME energy range (9-30\,keV) at the calibration facility in the Institute of High Energy Physics (IHEP). The double crystal monochromator was used to generate mono-energetic photons \citep{Cxl2019}. The fluorescence lines of Ag was found in the measured spectrum when the photon energies were above the K-edge of silver (25.5\,keV),  and we can use the measured spectrum and Monte Carlo (MC) simulation to confirm the thickness of Ag glue under the Si-PIN detectors of ME.

\begin{figure}[htb]
  \centering
  \subfigure[]{
     \includegraphics[width=0.22\textwidth]{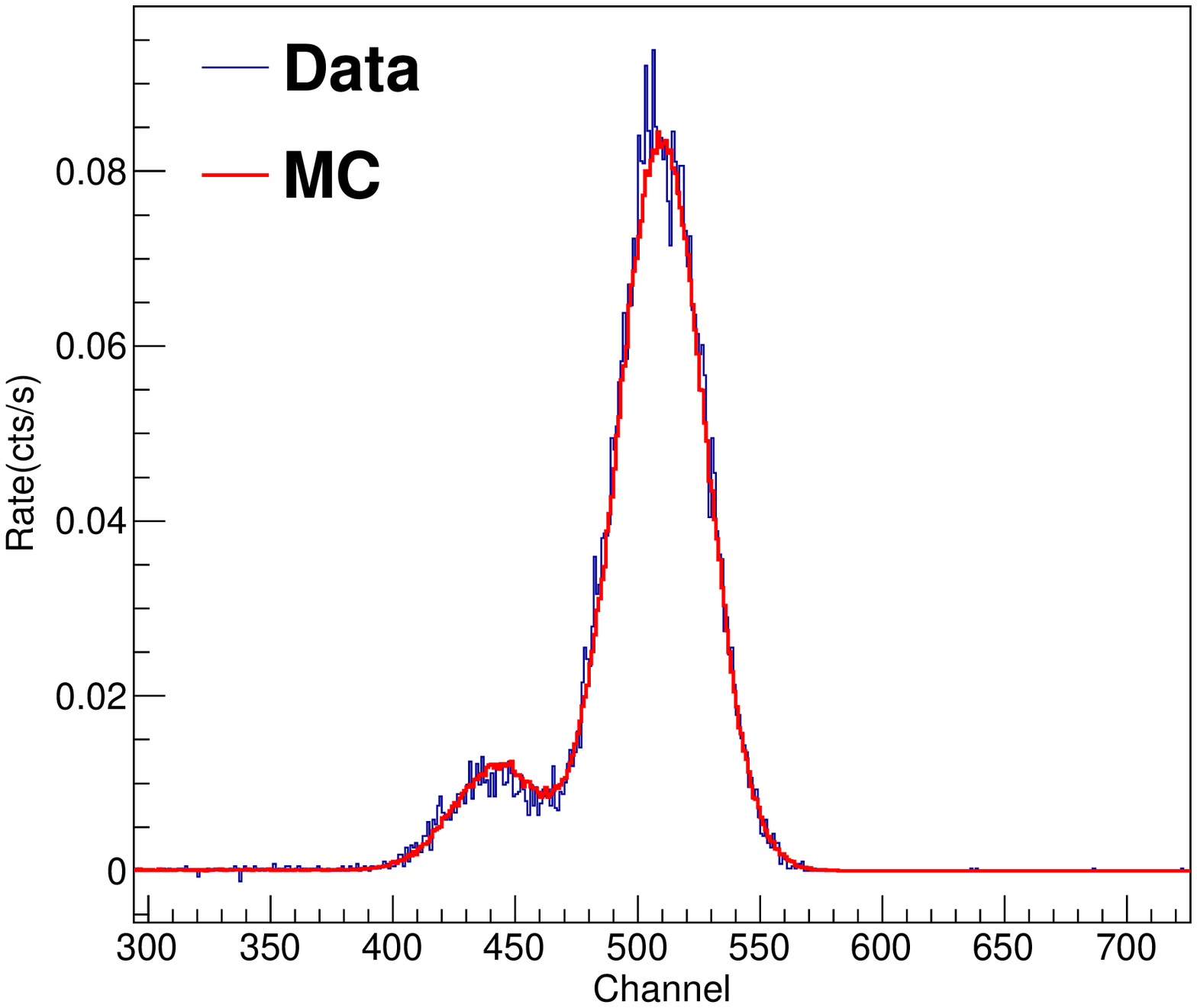}}
  \subfigure[]{
      \includegraphics[width=0.235\textwidth]{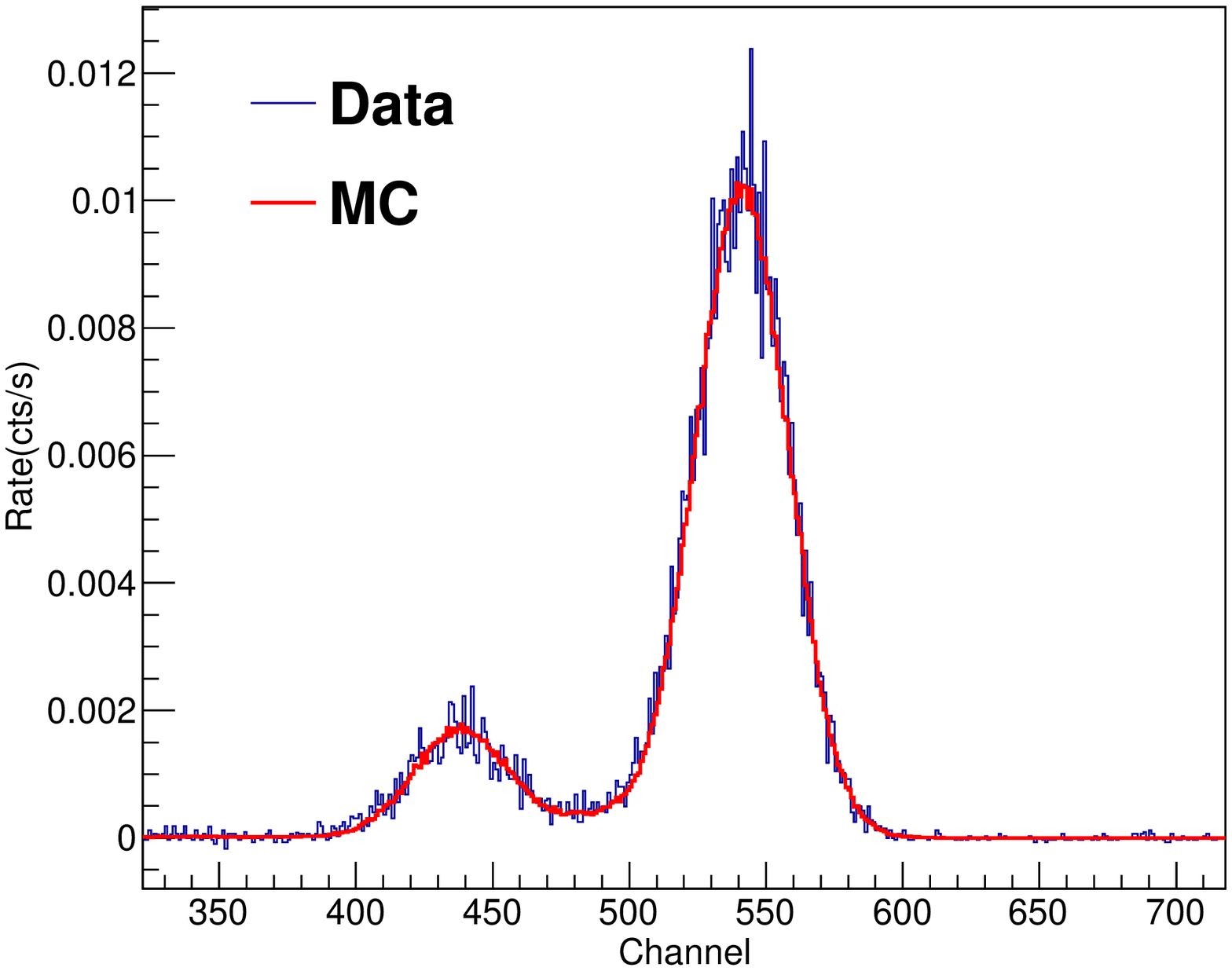}}
  \subfigure[]{
      \includegraphics[width=0.23\textwidth]{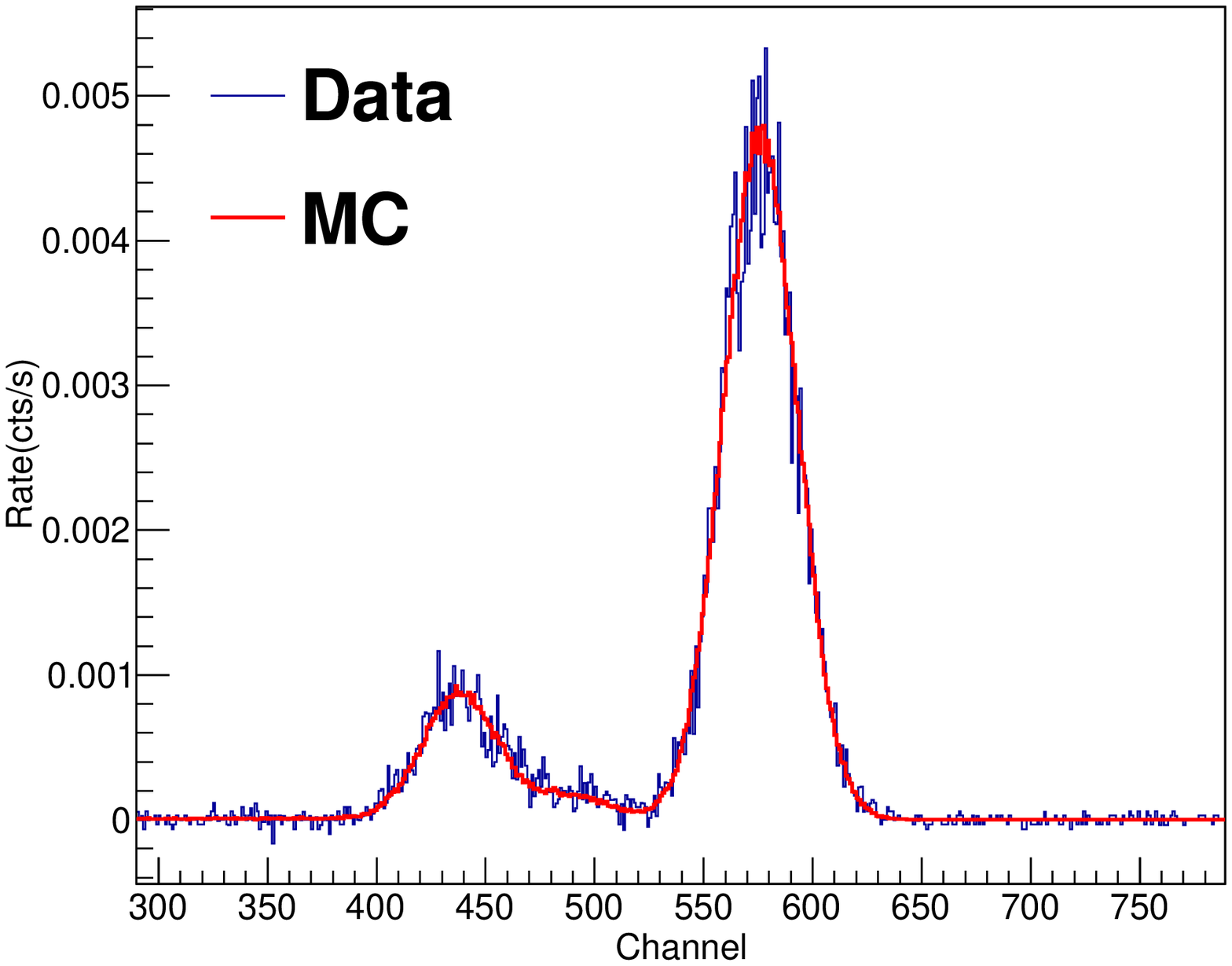}}
    \caption{(a) Simulated spectrum and measured spectrum with X-ray of 26\,keV. (b) Simulated spectrum and measured spectrum with X-ray of 28\,keV. (c) Simulated spectrum and measured spectrum with X-ray of 30\,keV. The thickness of Ag glue is estimated as 14\,um.}
  \label{FIG:16}
\end{figure}

We also utilize Geant4 (version 4.9.4) to perform the simulations. We invoke the low-energy electromagnetic process and consider the photoelectric effect, Compton scattering, Rayleigh scattering in the simulation. Fluorescence and Auger processes are also loaded in the photoelectric effect. The X-ray events spectrum resulting from monochochromatic radiation with an energy $E$ shows a simple Gaussian for photon energies below the Ag K-shell edge (25.5\,keV). But for photons with energy above the K-edge of Ag, it consists of a Gaussian photo-peak and Ag fluorescence peak at 22.5\,keV as displayed in Figure \ref{FIG:16}.
We have adjusted the thickness of Ag glue under the SI-PIN till the differences of MC simulation spectra and measured spectra are the minimum as shown in Figure \ref{FIG:16}. Finally, the thickness of Ag glue under the Si-PIN is estimated as 14\,um.

As the resolution of ME has changed only slightly in orbit as shown in Figure \ref{FIG:6_2}, so the redistribution matrix of ME is still taken as the pre-launch one.

\subsection{LE} \label{sec:4.3 rmf of LE}
As LE is a type of CCD detector, the spreading of the charge cloud over several pixels produced by a photon or a charged particle can be read out by several adjacent periods. To eliminate events due to charged particles, and to obtain good energy resolution, we only consider single events, which are not split as two or more adjacent readout, as valid X-ray events.

\begin{figure}
	\centering
\includegraphics[width=0.45\textwidth]{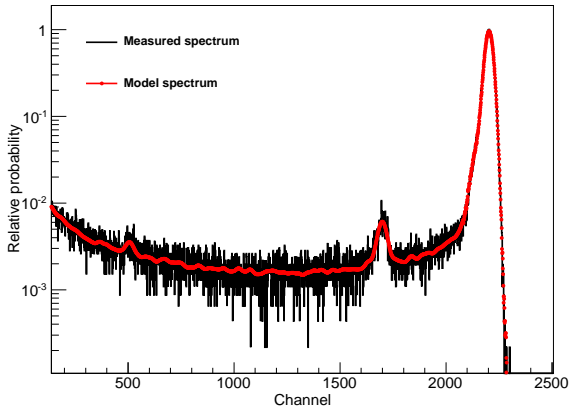}
	\caption{The measured spectrum of 7.48\,keV photons compared with the model. The model spectrum of 7.48\,keV is obtained by interpolating and smooth the spectra of photons with energies of 7.01\,keV and 8.02\,keV.}
	\label{FIG:17}
\end{figure}

The pre-launch RMF calibration of LE was done using the calibration facility in IHEP, using twenty discrete energies covering the 0.9-12\,keV energy range. The X-ray single events spectrum resulting from monochochromatic radiation with an energy $E$ significantly differs from a simple Gaussian. It consists of multiple components: a Gaussian photo-peak with a shoulder on the low energy side, a plateau extending to low energies. For photon energies above the Si K-shell edge (1.839\,keV) two additional features appear: an escape peak of energy $E-E_{\rm{Si}}$ and a Si $K_{\alpha}$ fluorescence peak at 1.74\,keV. Displayed in Figure \ref{FIG:17} is a measured spectrum with energy of 7.48\,keV.

At first, we also used Geant4 to simulate the response function of LE and considered the charge transfer process \citep{ChargeProcess} in the simulation. It is very difficult to use the same parameters, like the coefficient of the primary cloud radius, diffusion length in depletion region and so on to make all the MC spectra the same as the measured spectra.
Finally it was decided to use the two dimensional (one is the energy of incident photons and the other is the ADC channel of LE) interpolation to generate the response function of LE, since we have twenty discrete energies covering from 0.93\,keV to 11.9\,keV, and almost for every 0.5\,keV interval we have a measured spectrum below 8\,keV.
After interpolation, the probability in every ADC channel at energy $E$ will be calculated, and we smooth it to get the final model of response function.
As a verification, we compare the measured spectrum and model spectrum for photons with energy of 7.48\,keV and they show good agreement with each other as plotted in Figure \ref{FIG:17}. The model spectrum of 7.48\,keV is interpolated by photons with energy of 7.01\,keV and 8.02\,keV.
Therefore, the pre-launch RMF is generated from 0.855\,keV to 11.915\,keV and step is 0.01\,keV.

When we use the pre-launch RMF to fit the in-flight spectrum of Cas A, extra broadening is needed to fit the line profiles well as described in section \ref{sec:3.3 cal of LE resolution}. The extra broadening of the resolution is convolved with the pre-launch RMF to generate the RMF in-flight.

\section{Effective areas calibration}

The Crab is a center filled pulsar wind nebula powered by a pulsar with a period of 33\,ms. It has served as the primary calibration source for many hard X-ray instruments due to its brightness, relative stability, and simple power-law spectrum over the band from 1 to 100\,keV \citep{Kristin K}.  The Crab, however, is too bright for most CCD based focusing X-ray instruments because of the pile-up effect, and has been replaced with fainter sources, such as the Crab-like PWN G21.5+0.9\citep{Tsujimoto}.
As a collimated telescope, \emph{Insight-HXMT} has a high background level and does not suffer from pile-up, the Crab therefore remains the best choice for its calibration in the X-ray band covered by its three payloads.

The spectrum of the Crab in the 1--100\,keV X-ray band has been well-described by a power-law with photon index of $\Gamma \sim 2.11$ \citep{Kristin K}, \citep{E.Massaro}.
 As for the normalization factor, we have fitted the Crab spectrum measured by \emph{NuSTAR} in March, 2018 with $N = 8.76\,{\rm {keV^{-1}cm^{-2}s^{-1}}}$.
 The model of Crab as a simple absorbed power law is used as,
\begin{equation}\label{equ:crabmodel}
   F(E)=\rm{wabs}(\emph{E})\emph{N}\emph{E}^{-\Gamma},
\end{equation}
 where $E$ is the photon energy, $\rm{wabs}$ is the interstellar absorption, \emph{N} is the normalization factor and $\Gamma$ is the power-law photon index. We define the Crab model with $\Gamma=2.11$,\\
 ${N}=8.76\,{\rm {keV^{-1}cm^{-2}s^{-1}}}$ and ${N_{H}}=3.6\times10^{21}\,{\rm cm^{-2}}$ \citep{E.Massaro} .

After the background of the instruments is subtracted from the observed spectrum of Crab,
the net detected counts in a given instrumental pulse height bin, $S(Ch)$, can be derived according to the equation,
  \begin{equation}\label{equ:EffArea}
   S(Ch)=F(E)\times A(E)\ast RMF(Ch, E) ,
 \end{equation}
where $F(E)$ is the model photon spectrum of the Crab as a function of the incident photon energy, and $RMF(Ch, E)$ is the redistribution matrix, that represents the probability density in a given pulse height bin ($Ch$) for the photons with energy $E$.  $A(E)$ is the effective area, also known as the ancillary response function (ARF).

\begin{figure}
	\centering
\includegraphics[width=0.43\textwidth]{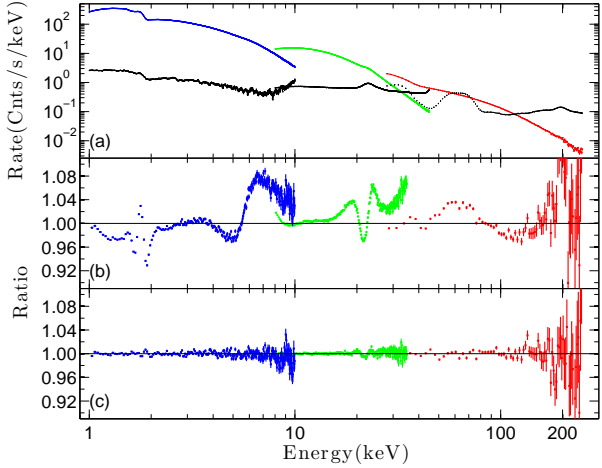}
	\caption{Panel (a) are the background-subtracted Crab spectra measured by LE (Blue), ME (Green) and HE (Red, summed over 17 un-blinded detectors) and the black is the corresponding estimated background for the three payloads.  The lower panels (b) and (c) show the ratio of the data to the Crab model ($\Gamma = 2.11$ , ${N}=8.76\,{\rm {keV^{-1}cm^{-2}s^{-1}}}$ , and ${N_{H}}=3.6\times10^{21}\,{\rm cm^{-2}}$) before and after the effective areas calibrated. }
	\label{FIG:18}
\end{figure}

Prior to launch, we used ground calibration results and Monte Carlo simulations based on Geant4 toolkit to produce the basic effective areas.
After launch, when we use the new resolution to simulate the effective areas, there still remain systematic residuals in the Crab spectrum. As plotted in panel (b) of Figure \ref{FIG:18}, it shows the ratio of the net data to the Crab model as defined in Equation (\ref{equ:EffArea}). The ratio shows several features that indicate the inaccuracies in the simulation of the effective areas and the estimation of the background. The background levels of the three instruments are also plotted in panel (a) of Figure \ref{FIG:18}.

 \begin{itemize}
  \item For LE, the response below the K-edge of Si shows deviations to the data because of the uncertainties in the thicknesses of materials in front of the depletion region, such as $\rm{SiO_{2}}$, $\rm{Si_{3}N_{4}}$ and Poly Si and so on. At around 1.8\,keV, the K-edge of Si causes the residuals due to the uncertainties of E-C or the resolution and some other factors. The residuals above 5\,keV are still unknown, and may be caused by the charge transfer process in the CCD, the thickness of the depletion region or the background estimation.
  \item For ME, the response around the Ag line at about 22 \,keV is due to the difference of Ag glue thickness and distributions under different Si-PIN detectors. The pre-launch response matrix calibration measured only one Si-PIN detector. The residuals above 25\,keV are caused by the estimation of background and also the thickness of Ag glue under the Si-PIN detectors.
  \item For HE, the pre-launch efficiencies of HE detectors with anti-coincidence detectors or without are well in agreement with the Monte Carlo simulations\citep{Lcz2019},\citep{XF.LiHE2019}.  The residuals of HE in-orbit are less than 4\% below 150\,keV and the shape is similar to the background spectrum. These may be caused by the uncertainties of HE background estimation.
\end{itemize}

Although it is desirable to have a completely physics-based effective areas, this is not usually achievable with limited calibration sources and time. Finally, it is decided to use an empirical function $f(E)$ to modify the simulated effective areas. Since $f(E)$ is a function of energy $E$, its effect should be folded through response matrix.
   This process can be summarized in the following steps:
 \begin{itemize}
  \item  We reduce all the Crab data in the two years observations using HXMTDAS V2.02 to generate the total spectrum (including Crab and background) and background spectrum. If the resolution is same in different observation data, we merge the total spectrum and background spectrum. For HE and ME, we have merged about two years observation data, but for LE, we have merged the Crab data in each month due to the evolution of the resolution.
  \item  According to the results of in-flight calibration, we reproduce $RMF(Ch, E)$ as described in Section \ref{RMF cal} and simulate effective areas $A(E)$ again.
   \item We make $f(E)$ multiply the Crab model $F(E)$ and convolve them with $RMF(Ch, E)$ and $A(E)$ generated in the second step to fit the merged background-subtracted spectra. The residuals of the Crab spectra can be derived as shown in panel (b) of Figure \ref{FIG:18}.
  \item We optimize the empirical function and make the residuals in an acceptable level as shown in panel (c) of Figure \ref{FIG:18}, The residuals are typically less than 2\% in most energy bands. The parameters of the empirical function can be derived and the effective areas in orbit can be represented as $f(E)*A(E)$.  The final residuals are less than several percent for the three instruments as shown in panel (c) of Figure \ref{FIG:18}.
\end{itemize}

 In panel (a) of Figure \ref{FIG:18}, we show the net Crab spectrum measured by \emph{Insight-HXMT}. For HE and ME we use all the Crab observations between September 2017 and April 2019, while for LE we only use the observations in November 2017.
From the bottom panel of Figure \ref{FIG:18}, the ratios of LE are less than $\pm 2\%$ up to 7\,keV.
Between 7 and 10\,keV, residuals are about $\pm 4\%$.
The ratios of ME are less than $\pm 1\%$ up to 20\,keV, and slightly higher above 20\,keV. There is also an artificial structure at around the characteristic X-ray energy of silver, suggesting that the response matrix needs to be improved.
The ratios of HE are less than  $\pm 2\%$ up to 120\,keV, but above 120\,keV, the deviations become larger, up to 4-10\%.

\begin{figure}
	\centering
\includegraphics[angle=-90,width=0.45\textwidth]{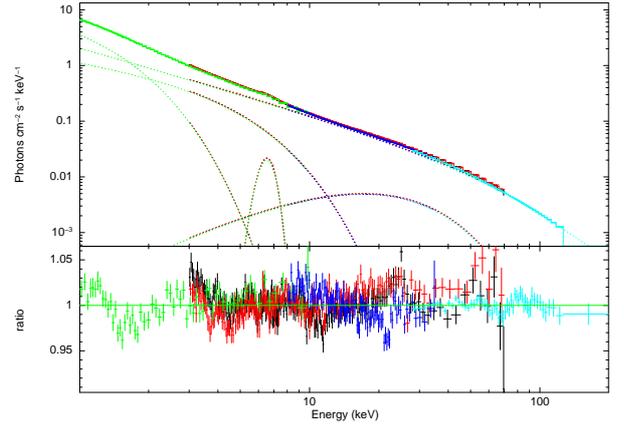}
	\caption{The background-subtracted spectrum of MAXI J1820+070 obtained by \emph{Insight-HXMT} and \emph{NuSTAR}, compared with the model (constant*(diskbb+diskbb+cutoffpl+cutofpl+gaussian)). The black and red lines and data points are the results from FPMA and FPMB of \emph{NuSTAR}. The green, blue and cyan are the results from LE, ME and HE of \emph{Insight-HXMT}, respectively. The bottom panel shows the data-to-model ratio of the five instruments. The constants for FPMA, FPMB, LE, ME and HE are 1, 1.02, 0.97, 0.98, and 0.95 respectively.}
	\label{FIG:19}
\end{figure}

After the in-flight effective areas are calibrated, some simultaneous observations with \emph{NuSTAR} like Swift J0243.6+6124, MAXI J1820+070 are used to validate the effective areas.
MAXI J1820+070 was observed with \emph{NuSTAR} simultaneously on March 24, 2018 while the flux was about 3\,Crab.
The observed data are processed using the standard pipelines of \emph{NuSTAR} and \emph{Insight-HXMT}.
The model is used as \\
constant*(diskbb+diskbb+cutoffpl+cutofpl+gaussian). \\
As shown in Figure \ref{FIG:19}, we simultaneously fit the spectrum of MAXI J1820+070 measured by \emph{NuSTAR} and \emph{Insight-HXMT} and allow the constant to float.
The differences of constant are within about 5\%  with respect to \emph{NuSTAR} and the residuals for the two telescope are also shown in Figure \ref{FIG:19}. Most of the ratios are within $\pm 3\%$.

\section{Systematic errors of the three instruments}

We have a lot of Crab monitoring observations with typical exposure time of about 12000\,s.
We reprocess all the Crab data with the new calibration files to verify if the effective areas are correct.
We fit individual observation of the Crab spectra and allow the parameters to vary. The parameters are the normalization factor, $N$, power law photon index, $\Gamma$, and the neutral absorption parameter $N_{H}$ which is fixed at $0.36\times10^{22}\,{\rm cm^{-2}}$. The energy fit range used for HE is 28--250\,keV. The energy fit band for ME is 10--35\,keV while for LE it is 1--10\,keV.
The systematic errors or scatter of normalization factor or photon index can be calculated by solving the equation,
\begin{gather}
    \sum_{i=1}^N \frac{(X_{i}- \overline {X})^{2}}{\sigma_{i}^{2}} = N-1,
    \label{equ_syserror}
\end{gather}
where
\begin{gather}
    \sigma_{i}^{2} = \sigma_{\rm sys}^{2} + \sigma_{{\rm stat},\,i}^{2},
     \label{equ_syserror3}
\end{gather}
\begin{gather}
    \overline{X} = \sum_{i=1}^N X_{i}\times w_{i}, \quad w_{i} = \frac{\frac{1}{\sigma_{i}^{2}}}{\sum_{i=1}^N \frac{1}{\sigma_{i}^{2}}}.
    \label{equ_syserror2}
\end{gather}

 Here $\sigma_{\rm sys}$ is the systematic error of $X$, $X_{i}$ is the fit parameters (such as normalization factor or power-law index) of each individual observation,  $\sigma_{{\rm stat},i}$ is the statistic errors of the fit parameters, and $\sigma_{i}$ is the total errors of the parameters. $N$ is the number of individual observations.

We define the bias of $X$ as $({\overline{X}-X_{\rm{model}}})/{X_{\rm{model}}}$
where $X_{\rm{model}}$ is the model values of Crab spectrum which are shown in Function (\ref{equ:crabmodel}).
The systematic scatter of $X$ is defined as $ {\sigma_{\rm sys}}/{\overline{X}} $.

\begin{figure}
	\centering
\includegraphics[width=0.45\textwidth]{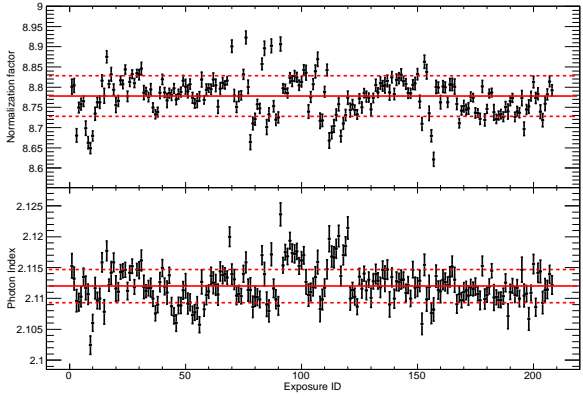}
	\caption{The normalization factor (top) and photon index (bottom) at different observation time of LE. The red lines are shown as the weighted mean value as defined in Equation \ref{equ_syserror2}. The red dashed lines mean the systematic error ($\pm 1\sigma$) of normalization and photon index.  }
	\label{FIG:20}
\end{figure}

The individual fit parameters are recorded and are used to compute the biases and systematic scatter of the normalization factor and power law index of the Crab spectrum. The results are shown in Table \ref{tab:tab4 biases}.
The results demonstrate that the normalization factor bias of HE is about 2.4\% for the observations with effective exposure time more than 1000\,s and increases with effective exposure time. The systematic scatter of HE normalization factor is about 3.5\% (1$\sigma$) and decreases with effective exposure time. The photon index bias of HE is small, 0.25\%, and the scatter is at about 0.5\% (1$\sigma$).
For ME, the normalization factor bias is about -1.2\% and also increases with exposure time  while the scatter is about 4.4\% (1$\sigma$) and decreases with time. The characteristic of the photon index is the same as the normalization factor but with smaller scatter.
For LE, the bias and scatter of the normalization factor and photon index are much smaller than that of HE and ME as shown in Table \ref{tab:tab4 biases}.
 The results are also shown graphically in Figure \ref{FIG:20} as an example of LE.

\begin{table}[htb]\footnotesize
  \centering
  \caption{The biases and scatter of Crab parameters for HE, ME and LE.  Normalization factor is $N$ and power-law photon index is $\Gamma$.}\label{tab:tab4 biases}
          \begin{tabular}{ccccccc}
    \hline
    \hline
      & Exposure & Observation & $N_{bias}$  & $N_{scatter}$  & $\Gamma_{bias}$ &$\Gamma_{scatter}$  \\
                 &     & times &  &     \\
    \hline
      & 500s-1000s &18 & 0.6\% & 5.2\% &0.003\% &  0.7\%\\
         & 1000s-2000s & 56 & 1.7\% &  4.3\% & 0.16\% &0.6\%  \\
     HE & 2000s-3000s& 79 &  2.4\%&  3.6\% &0.24\% &0.6\%\\
     & >3000s & 82 & 2.8\% & 2.9\% & 0.3\%& 0.4\% \\
     & >1000s & 217 & 2.4\% & 3.5\% & 0.25\%& 0.51\% \\
       \hline
     & 500s-1000s & 13 & -3.8\%& 8.0\% & -0.88\%& 1.53\% \\
     & 1000s-2000s & 74 & -2.0\% & 4.5\% &-0.40\% & 0.91\% \\
    ME    & 2000s-3000s & 128& -1.0\% & 4.4\% &-0.24\% & 0.84\%\\
        & >3000s & 41 & -0.08\% & 3.9\% &-0.07\% & 0.79\% \\
        & >1000s & 243 & -1.2\% & 4.4\% & -0.26\%& 0.86\% \\
       \hline
     & 500s-1000s & 22& 0.18\% & 0.75\% &  0.22\%& 0.14\% \\
     & 1000s-2000s & 102 & 0.21\%& 0.73\% & 0.12\% & 0.13\%  \\
     LE & 2000s-3000s & 83 & 0.17\%& 0.37\% & 0.08\% & 0.13\% \\
     & >3000s & 23 &0.33\% &0.36\%  & 0.11\% & 0.07\% \\
        & >1000s & 208 & 0.21\% & 0.57\% & 0.10\%& 0.13\% \\
   \hline
    \hline
    \end{tabular}
\end{table}

\begin{figure}
	\centering
\includegraphics[width=0.45\textwidth]{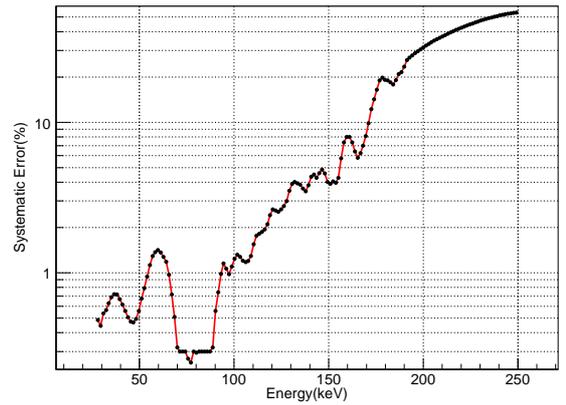}
	\caption{HE systematic errors of ratios (data to the Crab model) as a function of energy, which should be considered in the spectral fitting.}
	\label{FIG:21}
\end{figure}

\begin{figure}
	\centering
\includegraphics[width=0.45\textwidth]{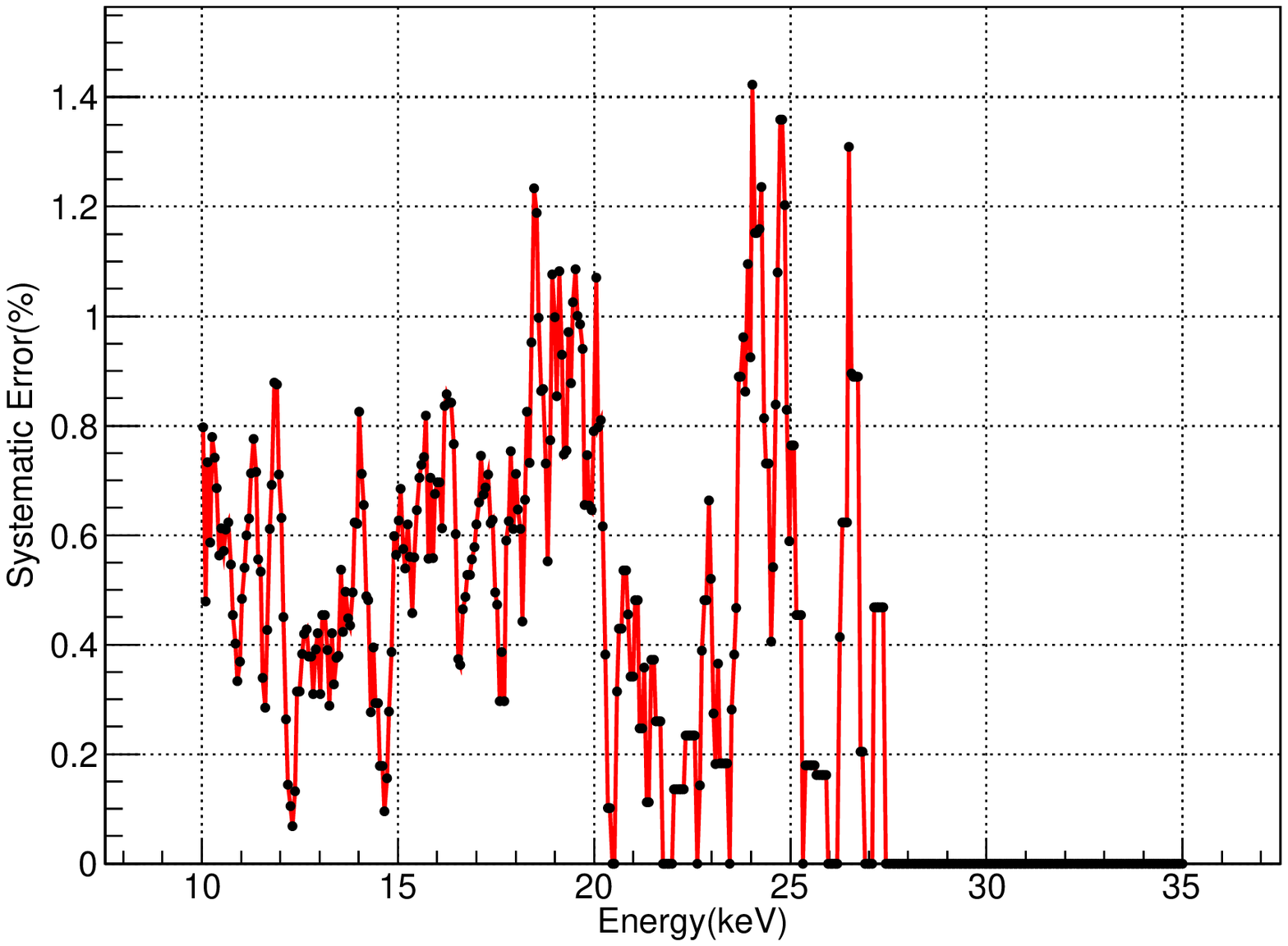}
	\caption{Same as Figure \ref {FIG:21}, but for ME.}
	\label{FIG:22}
\end{figure}

\begin{figure}
	\centering
\includegraphics[width=0.45\textwidth]{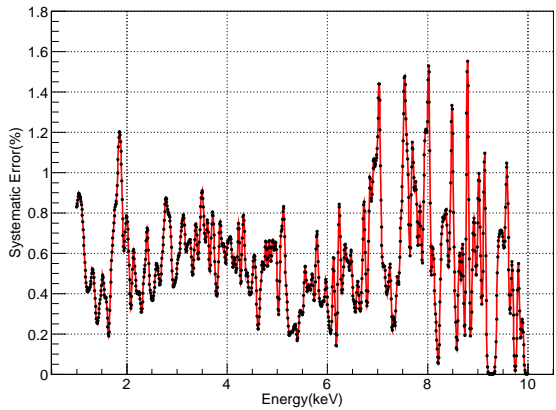}
	\caption{Same as Figure \ref {FIG:21}, but for LE.}
	\label{FIG:23}
\end{figure}

We have reprocessed all the Crab data and generated the corresponding response files for the three payloads at different observation time.
The model of Crab can be fixed to get the ratio (data to model) of each individual observation in each PI channel.
We can calculate the systematic errors and biases of the ratio at each PI channel using the same method as described in Equation (\ref{equ_syserror}). The biases of ratio in each PI channel are less than 1\% for HE (28--150\,keV),  ME (10-35\,keV) and LE (1-10\,keV).
The systematic errors of the ratios for the three \emph{Insight-HXMT} instruments are shown in Figures \ref {FIG:21}, \ref{FIG:22} and \ref{FIG:23}, respectively. These systematic errors of the ratios, compared to the model spectrum of the Crab nebular, can be used in the spectral fitting.

The systematic errors of HE in the spectral fitting are less than 2\% below 120\,keV. Between 120\,keV and 170\,keV, the systematic errors go up to 2\%-10\%. Above 170\,keV, the background count rate is about ten times higher than the rate of the Crab, and the systematic errors are dominated by the high background level and less photons from the Crab.

The systematic errors of ME in the spectral fitting are less than 1.5\% in 10--35\,keV band. From Figure \ref{FIG:22}, we can not obtain the systematic errors of ME above 27\,keV, because the errors of ME background are over estimated. We will re-estimate the error of ME background in the next background model.

The systematic errors of LE in the spectrum fitting are less than 1\% in 1--7\,keV except the Si K-edge at 1.839\,keV, up to 1.5\% and less than 2\% in 7--10\,keV.

\section{Conclusions}

\emph{Insight-HXMT} is a wide-band X-ray astronomical satellite with well calibrated instruments.
The energy gain and resolution of HE detectors are stable after three month in-flight and are determined with an accuracy of 1\%.
The in-orbit energy gain of ME shows a slow evolution at less than 1\% till now.
The uncertainty of the LE gain is less than $\sim 20 $ eV in 1--9\,keV band.
The effective areas are calibrated with the Crab nebular using a simple absorbed power law model when photon index $\Gamma =2.11$, normalization factor $N = 8.76\,{\rm {keV^{-1}cm^{-2}s^{-1}}}$ and interstellar absorption $N_{H} = 0.36\times10^{22}\,{\rm cm^{-2}}$.
The systematic error of HE in the spectral fitting is better than 2\% below 120\,keV and gradually increase toward higher energies, and that of ME is less than 1.5\%. As for LE, the systematic errors in the spectral fitting is less than 1\% in 1--7\,keV except the Si K-edge and slightly higher above 7\,keV.
We plan to continue improving the spectrum capability of the three instruments and monitor the detector gain, resolution and effective areas.

\section*{ }

   \textit{This work is supported by the National Natural Science Foundation of China under grants (No. U1838105, U1838201, U1838202) and the National Program on Key Research and Development Project (Grant No.2016YFA0400800). This work made use of data from the \emph{Insight-HXMT} mission, a project funded by China National Space Administration (CNSA) and the Chinese Academy of Sciences (CAS).}



%
%
%
%
\end{document}